\documentclass[review]{elsarticle}

\usepackage[a4paper, total={6.5in, 9.0in}]{geometry}

\usepackage{hyperref}
\usepackage{amsmath}
\usepackage{amssymb}
\usepackage{graphicx}% Include figure files
\usepackage{dcolumn}% Align table columns on decimal point
\usepackage{bm}
\usepackage{verbatim}

\usepackage{multirow}
\usepackage{float}

\usepackage[utf8]{inputenc}
\usepackage[T1]{fontenc}
\usepackage{mathptmx}
\usepackage{etoolbox}
\usepackage{color}

\journal{International Journal of Computational Fluid Dynamics}

\biboptions{authoryear}

\begin{document}

\begin{frontmatter}

\title{On the generalizability of machine-learning-assisted anisotropy mappings for predictive turbulence modelling}

%% authors
\author[mymainaddress]{Ryley McConkey \corref{mycorrespondingauthor}}
\cortext[mycorrespondingauthor]{Corresponding author: rmcconke@uwaterloo.ca}

\author[mymainaddress]{Eugene Yee}
\author[mymainaddress]{Fue-Sang Lien}

\address[mymainaddress]{Department of Mechanical and Mechatronics Engineering, University of Waterloo, 200 University Avenue, Waterloo, ON, N2L 3G1}

\begin{abstract}

Several machine learning frameworks for augmenting turbulence closure models have been recently proposed. However, the generalizability of an augmented turbulence model remains an open question. We investigate this question by systematically varying the training and test sets of several models. An optimal three-term tensor basis expansion is used to develop a model-agnostic data-driven turbulence closure approximation. Then, hyperparameter optimization was performed for a random forest, a neural network, and an eXtreme Gradient Boosting (XGBoost) model. We recommend XGBoost for data-driven turbulence closure modelling owing to its low-tuning cost and good performance. We also find that machine learning models generalize well to new parametric variations of flows seen in the training dataset, but lack generalizability to new flow types. This generalizability gap suggests that machine learning methods are most suited for developing specialized models for a given flow type, a problem often encountered in industrial applications.		
		
\end{abstract}


\end{frontmatter}

%\linenumbers

\section{\label{sec:1}Introduction}

Reynolds-averaged Navier-Stokes (RANS) simulations of turbulent flows are frequently used to help design industrial systems~\citep{Menter2001}. Turbulence modelling focuses on the approximation of the unclosed term in the RANS mean momentum transport equation: namely, the Reynolds stress tensor. Though eddy-viscosity based models are perhaps the lowest fidelity turbulence closure methods in common use, they are still the ``workhorse'' of industrial simulations~\citep{Witherden2017}. Higher-fidelity RANS closure methods such as Reynolds stress transport models (RSTM) are often bypassed in favour of large-eddy simulation (LES) due to various reasons such as numerical stability concerns. However, the major issue with LES is the high computational cost~\citep{CFD2030}. While technological advances help expand the speed and scale of available computational resources, the large difference in the computational cost between RANS and LES implies that RANS will continue to be used for the foreseeable future~\citep{Witherden2017}. Unfortunately, eddy-viscosity based RANS models have several well-known deficiencies that affect their accuracy~\citep{Pope2000,Wilcox1994,Layton2020}.

While the Reynolds stress tensor is the unclosed term in the RANS mean momentum equation, the anisotropic portion of this tensor is the portion responsible for momentum transport. Therefore, the majority of closure models are formulated in terms of the anisotropy tensor, rather than the Reynolds stress tensor. The isotropic component of the Reynolds stress tensor is related to the turbulent kinetic energy, and can be combined with pressure to form the modified pressure. The remaining anisotropic portion of the Reynolds stress tensor is of principal interest in closure modelling, since it is the unclosed influence of turbulence on mean momentum balance.

A promising method to overcome the deficiencies in RANS modelling is the application of machine learning to model the turbulence closure relationships~\citep{Duraisamy2019,Brunton2020}. Though these closure relationships have been the focus of decades of RANS turbulence modelling, they remain the principal weakness in industrial RANS. Recent advances in machine learning~\citep{Ling2016,Wu2018,Kaandorp2020} and the recent availability of open-source turbulence datasets have accelerated the progress in this area~\citep{JHTB, NASAturbmodel, ERCOFTAC, McConkeySciDataPaper2021}. Machine learning has also been used to augment RSTM~\citep{Panda2021,Panda2021a} as well as subgrid-scale modelling in LES~\citep{Maulik2019}. Within the scope of machine learning augmented eddy-viscosity modelling, a number of open questions remain. How should the model predictions be injected into the RANS equations? How well do these models generalize? What machine learning algorithm is best suited for this problem? Investigations in this area have overwhelmingly focused on the first question of injecting the machine-learnt Reynolds stress tensor into the RANS framework---this question is undoubtedly the most important starting point. However, few studies have addressed the latter two questions. Perhaps the next most important question that needs to be addressed is the generalizability of data-driven turbulence closure models.

In this study, we evaluate models from three classes of machine learning algorithms for regression on tabular data: namely, neural network models, random forest models, and gradient boosting models---more specifically, eXtreme Gradient Boosting (XGBoost)~\citep{xgboostChen2016} models. To compare these models, we propose a new model-agnostic data-driven turbulence closure framework. We also perform a systematic generalizability investigation where we vary the flows contained in the training dataset. Our main findings are to recommend XGBoost for data-driven turbulence closure modelling, and to demonstrate a generalizability gap that exists within machine learning turbulence closures. This work is limited to an {\it a priori} investigation (i.e., no machine-learnt predictions of the Reynolds stresses are injected into the RANS model) in order to allow for a more in-depth analysis of the generalizability and algorithm selection.

The model comparison investigation in the present work mirrors typical machine learning studies where several models are compared on an identical problem. These types of studies are not common within data-driven turbulence modelling, as most studies are focused on presenting new holistic injection frameworks, and the majority of effort is therefore invested in implementing and testing the injected fields. The machine learning algorithm used for the turbulence closure mapping is almost always chosen to be a neural network or random forest~\citep{Ling2016,McConkey2022,Song2019,Yin2020a}, despite the machine learning community recommending gradient boosted tree ensembles (e.g., XGBoost, CatBoost, LightGBM) for tabular data regression~\citep{Shwartz-Ziv2022}. In terms of model structure, our framework aims to produce a model-agnostic tensor basis closure framework. For the present work, this allows us to compare various algorithms for predicting the basis tensor coefficients. However, decoupling the machine learning algorithm from the turbulence closure approximation also allows us to take advantage of new tabular data algorithms. The rapid pace of machine learning innovation means that new tabular data algorithms are likely to be developed, and a model-agnostic closure framework allows modular "swapping" of the machine learning algorithm itself. Other augmented closure frameworks such as the tensor basis neural network (TBNN)~\citep{Ling2016} and the tensor basis random forest (TBRF)~\citep{Kaandorp2020} are not model-agnostic, and require more adaptation as machine learning algorithms evolve.

This paper is structured as follows: Section~\ref{sec:methodology} describes the machine learning methodology including introducing the closure framework (\ref{sec:tensorbasis}), the machine learning algorithms (\ref{sec:models}), input features (\ref{sec:features}), dataset (\ref{sec:dataset}), hyperparameter optimization procedure (\ref{sec:hyperparameter}), and code (\ref{sec:libraries}). Section~\ref{sec:results} first presents the calculated optimal basis coefficients (\ref{sec:coefficients})  and then summarizes the machine learning model predictions for these coefficients (\ref{sec:mlresults}). The generalizability investigation is presented in Section~\ref{sec:generalizability}, with additional results in the supplementary information files. Conclusions are given in Section~\ref{sec:conclusion}.

\section{\label{sec:methodology}Methodology}
The RANS continuity and mean momentum equations are given by Equations~(\ref{eq:continuity}) and (\ref{eq:momentum}):
\begin{align}
	\nabla \cdot \vec{U} &= 0\ ,\label{eq:continuity}\\
	\nabla\cdot (\vec{U}\vec{U}) &= - \nabla p + \nu \nabla^2 \vec{U} - \nabla \cdot \tau\ ,\label{eq:momentum}
\end{align}
where $\vec{U} \equiv (U,V,W)$ is the steady-state mean velocity field, $\nu$ is the molecular kinematic viscosity of the fluid, and $p$ is the pressure. Here, the unclosed term $\tau$ is the Reynolds stress tensor. Predicting the Reynolds stress tensor as a function of the other mean flow variables in Equation~(\ref{eq:momentum}) is the turbulence closure problem. Often, the Reynolds stress tensor is decomposed into an isotropic ($(2k/3)\delta_{ij}$) and an anisotropic ($a$) component where $k$ is the turbulent kinetic energy (TKE) and $\delta_{ij}$ is the Kronecker delta function. Then, the mean momentum equation is written as
\begin{equation}
	\nabla\cdot (\vec{U}\vec{U}) = - \nabla \left(p+\frac{2}{3}k\right) + \nu \nabla^2 \vec{U} - \nabla \cdot a\ ,\label{eq:momentum_modified}
\end{equation}
where $p' \equiv (p+2k/3)$ is the modified pressure and $a$ is the dimensional anisotropy tensor. Moreover, $a$ can be non-dimensionalized as $b = a/(2k)$, where $b$ the non-dimensional anisotropy tensor (referred to simply as the anisotropy tensor in the present work). In view of this, the mean momentum equation is given by 
\begin{equation}
	\nabla\cdot (\vec{U}\vec{U}) = - \nabla \left(p+\frac{2}{3}k\right) + \nu \nabla^2 \vec{U} - \nabla \cdot (2kb)\ .\label{eq:momentum_modified2}
\end{equation}
The turbulence closure problem is now focused on the prediction of $b$ and $k$. For the majority of data-driven closure modelling, $k$ is modelled using the familiar transport and scale-determining equation, while $b$ is the focus of a complex closure relationship. Section~\ref{sec:machinelearning} provides more details concerning the modelling of $b$.

\subsection{\label{sec:machinelearning}Machine learning for RANS closure modelling}
Several recent investigations have presented various ways to augment RANS simulations using machine learning~\citep{Wang2017,Ling2016,Wu2018,Kaandorp2020,Liu2021}. In the present work, we focus on generalized eddy-viscosity modelling. A seminal work in this field was the introduction of the tensor basis neural network (TBNN) by \cite{Ling2016}. The goal of the TBNN is to predict the anisotropy tensor $b$ using the general ten-term tensor basis expansion originally proposed by \cite{Pope1975}:
\begin{equation}\label{eq:tentensorbasis}
	b = \sum_{n=1}^{10} g^{(n)}T^{(n)}\ ,
\end{equation}
where $g^{(n)}$ are scalar coefficients for the following basis tensors $T^{(n)}$ ($n=1,2,\ldots,10$):
\begin{align*}
	&T^{(1)} =S\ , && T^{(6)}= R^2 S + SR^2 - \tfrac{2}{3}I\text{tr}(SR^2)\ ,\\
	&T^{(2)} =SR - RS\ , && T^{(7)} =  RSR^2 - R^2 S R\ ,\\
	&T^{(3)} =S^2 - \tfrac{1}{3}I \text{tr}(S^2)\ , && T^{(8)} = SRS^2 - S^2 RS\ ,  \\
	&T^{(4)} =R^2 - \tfrac{1}{3}I \text{tr}(R^2)\ ,  && T^{(9)} = R^2 S^2 + S^2R^2 - \tfrac{2}{3}I\text{tr}(S^2R^2)\ ,\\
	&T^{(5)} =RS^2 - S^2R\ , && T^{(10)} = RS^2 R^2 - R^2S^2 R\ .
\end{align*}
Here, $I$ is the identity tensor, ${\rm tr}(\ \cdot \ )$ is the trace operator, and the basis tensors are constructed from the non-dimensional mean strain- and rotation-rate tensors defined, respectively, as follows:
\begin{equation}\label{eq:normalizedstrainrate}
	S \equiv \frac{T_t}{2}\left(\nabla U + \nabla U^\text{T}\right)\ ,
\end{equation}
and
\begin{equation}\label{eq:normalizedrotationrate}
	R \equiv \frac{T_t}{2}\left(\nabla U - \nabla U^\text{T}\right)\ ,
\end{equation}
where $T_t$ is a turbulent time scale (cf.~Equation~(\ref{eq:T_t})) and the superscript $\text{T}$ denotes tensor transposition.

In \cite{Ling2016}, the machine learning algorithm used was a deep neural network model. This algorithm was also used in subsequent TBNN-based investigations conducted by \cite{Song2019} and by \cite{Zhang2019a}. However, more recent investigations have often relied on random forest predictors. \cite{Kaandorp2020} developed the tensor basis random forest (TBRF) which is essentially the random forest analogue to the TBNN proposed by \cite{Ling2016}. Other notable work in augmented eddy-viscosity modelling includes the investigation conducted by \cite{Wu2018} which uses a model that predicts the components of an eigendecomposition of the anisotropy tensor. \cite{McConkey2022} decomposed the anisotropy tensor into optimal linear and nonlinear components. Two neural networks (one per component) were then used to learn the turbulence closure relationship for the anisotropy tensor on a periodic hills dataset.

In parallel with frameworks for predicting various forms of the anisotropy tensor $b$, \cite{Cruz2019} proposed the prediction of the Reynolds force vector (RFV) $\nabla \cdot \tau$ rather than $\tau$ itself. Since the divergence of the Reynolds stress tensor (viz., the vector $\nabla\cdot\tau$) appears in the mean momentum equation, the closure model can be simplified if this vector is predicted directly. This framework avoids propagating any errors associated with $\tau$ through the divergence operator, and has been demonstrated to be another promising way to augment RANS simulations. Recently, \cite{Brener2022} further improved this strategy, showing that it is a promising approach for data-driven closure modelling. In terms of the machine learning algorithm, \cite{Brener2022} chose to use a random forest methodology. However, a disadvantage of the RFV approach is that the training dataset needs to contain various derivatives of the Reynolds stress tensor, quantities which are often not included in direct numerical simulation (DNS) datasets.

Aside from the closure relationship itself, which has varied significantly between investigations, the injection procedure is another key ingredient for a data-driven closure. Perhaps the most important issue for injection is the conditioning problem. If the RANS equations are ill-conditioned, errors in the predicted turbulence closure term can be amplified and result in large errors in the predicted velocity field. While originally thought to be solved by using an implicit treatment of the eddy viscosity~\citep{Wu2019}, \cite{Brener2021} showed recently that the conditioning problem can be resolved by incorporation of the information concerning the high-fidelity velocity field. As this finding is relatively recent, few injection procedures currently make use of this key result~\citep{McConkey2021c,McConkey2022,Brener2022}. Though injection into the momentum equation is not considered in the present work, we base our closure and machine learning procedure on a one-time correction (open-loop) closure framework~\citep{Ho2021}. In an open-loop closure framework, the machine learning model is called to correct the closure term based on a converged RANS solution. Then, these fields are injected and fixed in the RANS equations until a new, ideally more accurate solution for the velocity field is obtained. Most previous work in augmented RANS modelling has focused on one-time correction. Closed-loop frameworks have also been proposed~\citep{Liu2021}, though the machine learning procedure and injection process varies for each of these frameworks. The key contributions of the present paper (namely, algorithm selection and generalizability) often receive less attention than the injection formulation, despite their importance to the long-term goals of data-driven turbulence modelling. 

\subsection{\label{sec:tensorbasis}Optimal tensor basis expansion for the Reynolds stress anisotropy}

The core assumption of the eddy-viscosity paradigm is that the anisotropy tensor is related to the mean velocity gradients~\citep{Pope2000}. The most common linear-eddy viscosity model (LEVM) is formulated as

\begin{equation}
	b = g^{(1)}T^{(1)} = -C_\mu S\ ,
\end{equation}
where $C_\mu$ is a global constant (closure) coefficient (usually, $C_\mu=0.09$).

However, other relationships have been proposed. These include nonlinear eddy-viscosity models (NLEVM) involving higher-order terms of the mean strain and rotation rate tensors. Motivated by deficiencies in the LEVM approximation, several NLEVM models have been previously proposed (e.g., \cite{Craft1996}, \cite{Lien1996}, \cite{Shih1995}), with most NLEVM models containing quadratic or cubic terms. In traditional turbulence modelling, the inclusion of additional terms means tuning additional (closure) coefficients, a process that becomes increasingly difficult as the number of terms increases. \cite{Pope1975} showed that a complete basis for three-dimensional flows involves ten tensors, up to fifth order in terms of strain- and rotation-rate tensors. A fifth-order eddy-viscosity model with ten closure coefficients $g^{(n)}$ ($n=1,2,\ldots,10$) would be insuperably difficult to tune using the conventional turbulence modelling paradigm. 

The investigation conducted by \cite{Ling2016} revitalized the turbulence closure modelling by using a neural network to predict the ten coefficients in the ten-term ``general nonlinear eddy-viscosity model'' proposed by \cite{Pope1975}. This ten-term model has been used widely in frameworks with a tensor basis expansion at the core of the closure relationship.

While the ten-term tensor basis used in several previous studies~\citep{Ling2016, Kaandorp2020, Song2019}) forms a complete basis for the anisotropy tensor for three-dimensional turbulent flows, a basis with fewer terms is sufficient to form a complete basis in two-dimensional flows. \cite{Gatski2000} showed that a tensor basis with just three terms is sufficient to model exactly the anisotropy tensor for two-dimensional (2D) flows, and this three-term basis can be interpreted as a least-squares approximation of the complete ten-term basis form for the anisotropy tensor associated with general three-dimensional (3D) flows~\citep{Gatski2000}. All of the classical nonlinear eddy-viscosity models contain less than ten terms, making them simpler than the general case and easier to tune. However, the three-term basis has received little attention in machine learning augmented turbulence closure modelling, despite the simplicity and theoretical justification for it, and despite the fact that the majority of investigations have focused on two-dimensional flows (which only require a three-term basis). We therefore use a three-term tensor basis for the anisotropy tensor. This tensor basis expansion for $b$ assumes the following form:

\begin{align}\label{eq:b}
	b &= g^{(1)}T^{(1)}+g^{(2)}T^{(2)}+g^{(3)}T^{(3)}\\
	&= g^{(1)}S + g^{(2)}(SR-RS) + g^{(3)} (S^2-\tfrac{1}{3}I\text{tr}(S^2))\ .
\end{align}

Using such a basis, the closure problem simplifies to the determination (specification) of the basis tensor coefficients $g^{(n)}$ in Equation~(\ref{eq:b}). By far, the most common approach is the linear-eddy viscosity model. Within this framework, a linear-eddy viscosity model is generated by specifying a global closure constant $g^{(1)} = - C_\mu$ (typically, $C_\mu=0.09$), and setting $g^{(2)}=g^{(3)}=0$. However, there are two principal deficiencies with this approach: namely, a strictly linear relation between the anisotropy tensor and the mean strain-rate tensor often breaks down~\citep{Pope2000}, and the optimal values of the $g^{(n)}$ coefficients are functions of the spatial coordinates---so, assuming that these coefficients are global constants introduces predictive inaccuracy. The latter disadvantage applies to both linear models (where techniques such as damping functions help adjust $C_\mu$ near the wall) and several nonlinear models \citep{Shih1995, Lien1996, Craft1996}.

To allow both a spatially-varying set of coefficients $g^{(n)}$ and a nonlinear representation for $b$ in terms of the mean velocity gradients, we propose an \textit{optimal} tensor basis expansion. In this expansion, the optimal coefficients $g^{(n)}$ are the solution of the following constrained multiple objective least-squares problem:

\begin{equation}\label{eq:optimization_problem}
	\begin{aligned}
		\min_{g^{(n)}} & \Bigl(||b_\theta - \sum_{n}g^{(n)}T^{(n)}||^2 + \lambda||g^{(n)}||^2\Bigr)\ ,\quad {\rm subject\ to}\\
		& g^{(1)} \leq 0\ ,\\
		& -\tfrac{1}{3} \leq \sum_n g^{(n)}T^{(n)}_{ii}\leq \tfrac{2}{3}\quad i=1,2,3\ , \\
		& -\tfrac{1}{2} \leq \sum_n g^{(n)}T^{(n)}_{ij}\leq \tfrac{1}{2},\quad i,j=1,2,3;\ i\neq j\ ,\\
	\end{aligned}
\end{equation}
for $n=1,2,3$, $\|\ \cdot\ \|$ denotes the Euclidean (L2) norm, the subscript $\theta$ indicates a quantity obtained from either DNS or LES, and $\lambda$ is a tuning (regularization) parameter that adjusts the relative importance of each component of the objective function. Based on a trade-off study, we used $\lambda = 1 \times 10^{-5}$, which provided a good balance between an accurate approximation of $b_\theta$  and well-behaved coefficients. 

It is noted that the formulation in Equation~(\ref{eq:optimization_problem}) can also be extended to an arbitrary tensor basis.  In words, the formulation in Equation~(\ref{eq:optimization_problem}) seeks to obtain the solution to the following optimization problem: namely, find the solution that represents the best compromise between minimizing the errors in the approximation of the anisotropy tensor while constraining the magnitude of the coefficients $g^{(n)}$ to be as small as possible (viz., shrinking the values of these coefficients toward zero). In particular, the objective function includes an L2 regularization (penalty) term
that constrains the norm squared of the optimized coefficients $g^{(n)}$. We found this regularization term was crucial for stabilizing and reducing the magnitude of these coefficients, thereby increasing their useability as machine learning prediction targets. 

The following constraints are also applied in Equation~(\ref{eq:optimization_problem}): namely, (1) ensure a non-negative eddy viscosity for simulation stability and (2) ensure a realizable anisotropy tensor through the application of realizability constraints on the various components of $b$. The first constraint is motivated by the fact that the term $g^{(1)}T^{(1)} = -C_\mu S$ is usually treated implicitly in a RANS  simulation. This implicit treatment is achieved by the use of an ``effective'' viscosity, $\nu_\text{eff} = \nu + \nu_t$, where $\nu$ is the molecular kinematic viscosity and $\nu_t$ is the eddy (turbulent) viscosity. Using the common turbulent timescale $T_t= k / \varepsilon$, we have $\nu_t = C_\mu \frac{k^2}{\varepsilon}$. The constraint $g^{(1)} \leq 0$ guarantees $\nu_t\geq 0$ (assuming $k$ is positive), an important target for stability in a RANS simulation. The remaining constraints arise from bounds on the anisotropy tensor itself in order to guarantee realizability~\citep{Banerjee2007}.

The optimal tensor basis expansion shown here is also invariant. The invariance of such tensor basis expansions motivated the use of the tensor basis neural network (TBNN) in \cite{Ling2016} and the subsequent application of the tensor basis random forest (TBRF) by \cite{Kaandorp2020}. However, our framework differs in one major way from these previous tensor basis models: namely, the optimal coefficients are calculated before training, and the model is trained to predict these coefficients directly. In principle, this simplifies the machine learning procedure by isolating the problem of optimizing $g^{(n)}$ and the model prediction of these optimal coefficients via the input features. This optimization problem is treated implicitly within the machine learning procedure of the TBNN and TBRF---therefore, control and insight into this optimization problem is lost. Further investigation will be required to determine the advantages of isolating this optimization problem from the machine learning process.

In the calculation of $g^{(n)}$ in Equation~(\ref{eq:optimization_problem}), the only quantity that comes from DNS or LES is the anisotropy tensor $b_\theta$. Therefore, the optimal $g^{(n)}$ coefficients approximate $b_\theta$ using $T^{(n)}$ from RANS, because only $T^{(n)}$ from RANS will be available at injection time. An alternative method to calculate the optimal $g^{(n)}$ coefficients would be to solve Equation~(\ref{eq:optimization_problem}) using $T^{(n)}_\theta$ (DNS or LES velocity gradients). However at injection time, the error $\epsilon = T^{(n)}_\theta - T^{(n)}$ will then be propagated and result potentially in larger errors in $b$. Therefore, using $T^{(n)}$ from RANS in Equation~(\ref{eq:optimization_problem}) embeds an implicit error-correction into the optimal $g^{(n)}$ coefficients, so that we do not rely on the implicit assumption that $T^{(n)}\approx T^{(n)}_\theta$ when the data-driven turbulence closure model is injected into the RANS model.

\subsection{Machine learning models}\label{sec:models}
A number of machine learning models were used in the present work to predict the optimal tensor basis coefficients $g^{(1)}$, $g^{(2)}$, and $g^{(3)}$. As discussed in Section~\ref{sec:tensorbasis}, these scalars were chosen as machine learning targets due to their optimal and realizable representation of the anisotropy tensor, and the embedded invariance of this representation. 

Previous data-driven closure investigations have used a variety of machine learning model types. The usual data-driven turbulence closure modelling problem is an example of a ``tabular data'' regression problem.  Within the turbulence modelling community, the majority of studies utilize neural networks and random forests for solving this tabular data problem. However, recent investigations in the broader machine learning field have revealed that neural networks are often not the optimal model type for tabular data. While neural networks excel at complex tasks such as computer vision and language processing, they are often outperformed by gradient boosting models on tabular data~\citep{Shwartz-Ziv2022}. Specifically, XGBoost is a popular tree-based gradient boosting framework for tabular data~\citep{xgboostChen2016}. XGBoost is an ensemble learning method in which a decision tree ensemble is created. As implied by the name, this ensemble is created by gradient boosting techniques. In gradient boosting, weak learners (in this case, decision trees) are repeatedly added to the ensemble to reduce the prediction errors of previous learners. Due to their ease of training and effective ensemble predictions, gradient boosting methods are highly popular for tabular data~\citep{Shwartz-Ziv2022}.

Despite the widespread use of XGBoost (and other similar implementations of gradient boosting machines such as LightGBM~\citep{lightgbmKe2017} and CatBoost~\citep{catboostProkhorenkova2018}) for tabular data within the broader machine learning community, these models are not popular within the turbulence modelling community. To explore viability for data-driven closure modelling, we include XGBoost as a model candidate for predicting the optimal tensor basis coefficients. In addition to XGBoost, we consider also a deep feedforward neural network (multi-layer perceptron or MLP) and a random forest model---the two main model types currently in use for data-driven turbulence closure modelling. Finally, we consider a simple super-ensemble of these three model types using two weighting schemes: namely, an equal weighting and a score-based weighting. The model types evaluated in the present study are summarized in Table~\ref{tbl:modeltypes}.

After training these three model types, two forms of stacking were used to combine the individual model predictions. This stacked ensemble consisted of three models (viz., the optimized or fully-trained version of each model type summarized in Table~\ref{tbl:modeltypes}). The ensemble predictions were calculated using:

\begin{equation}
	g^{(n)}_\text{ens} = \sum_{i=1}^N w_i g^{(n)}_i\ ,\qquad n=1,2,3\ ,
\end{equation}
where $w_i$ are the ensemble weights. Two weight calculation techniques were used. The first was equal weighting of each ensemble member, so $w_i=1/N$ where $N$ is the number of members (models) in the ensemble ($N=3$ in the present study). The second was a cross-validation score-based weight of the three models. In this case, the weights were calculated as follows:

\begin{equation}
	w_i = \dfrac{N-1}{N} - \dfrac{\text{C}_i}{\sum_{i=1}^N \text{C}_i}\ ,
\end{equation}
where $\text{C}_i$ is the cross-validation score for model $i$ ($i=1,2,\ldots,N$ with $N=3$).

These two ensembles are considered ``simple stacks''. Further investigation is required to determine the feasibility and advantages of using deep stacking (where the weights are determined locally by a more sophisticated machine learning algorithm which itself could be a gradient boosting machine) for turbulence closure modelling.

\subsection{\label{sec:features}Input feature selection}

As discussed in Section~\ref{sec:tensorbasis}, careful selection of the machine learning targets guarantees framework invariance and stability. Invariance and stability of the input features used is equally important. If both the target quantities and input features are invariant, the overall framework is invariant (under coordinate transformations and Galilean boosts).

To provide a fair comparison between the various model types, all models used the same input feature set. This input feature set consisted of a number of invariant scalars, combined with a set of heuristic scalars. The procedure for generating the invariant scalar set was discussed in detail in~\cite{McConkey2022}. In summary, these invariant scalars arise from the first and second invariants of a set of basis tensors formed by combining the mean strain rate, mean rotation rate, turbulent kinetic energy gradient, and pressure gradient. As discussed in Section~\ref{sec:dataset}, the present study uses the $k$-$\omega$ (shear stress transport) SST model as the base RANS turbulence closure model. Therefore, the pressure gradient tensor was used in the present work, rather than the $v^2$ (or, wall-normal stress) gradient tensor used in~\cite{McConkey2022}. Invariant scalars that were shown to be identically zero in \cite{McConkey2022} for two-dimensional flows were eliminated, as all of the flows used in the present study are two-dimensional in nature.

The heuristic scalars were selected from a set of scalars used by \cite{Kaandorp2020}. These scalars included the ratio of excess rotation to strain rate ($q_1$), the wall-distance based Reynolds number ($q_2$), and the ratio of turbulent time scale to mean strain timescale ($q_3$). These three heuristic scalars are defined as follows:
\begin{align}
	q_1 &= \frac{||R^2|| - ||S^2||}{2||S^2||}\ , \label{eq:q1}\\
	q_2 &= \text{min}\left(\frac{\sqrt{k} y_w}{50 \nu},2\right)\ , \label{eq:q2}\\
	q_3 &= \frac{k}{\varepsilon} ||S||\ , \label{eq:q3}
\end{align}
where $y_w$ is the wall-normal distance, $\varepsilon$ is the turbulent kinetic energy dissipation rate, and $\|\ T \|$ is the Frobenius norm of a tensor $T$. Table~\ref{tbl:inputfeatures} summarizes the input feature set used.

As described in Section~\ref{sec:dataset}, the base turbulence closure model for this study was the $k$-$\omega$ SST model. The turbulent timescale used to non-dimensionalize the strain- and rotation-rate tensors is determined from
\begin{equation}\label{eq:T_t}
	T_t = \text{max}\left( 6\sqrt{\frac{\nu}{\varepsilon}}, \frac{k}{\varepsilon} \right)\ .
\end{equation}
This timescale appears in the invariants that correspond to the strain- and rotation-rate tensors in Table~\ref{tbl:inputfeatures} (cf.~Equations~(\ref{eq:normalizedstrainrate}) and (\ref{eq:normalizedrotationrate})). The first argument in Equation~(\ref{eq:T_t}) (Kolmogorov timescale) was added to ensure the stability of $T_t$ near the wall. The TKE dissipation rate $\varepsilon$ was calculated using
\begin{equation}
	\varepsilon = \beta^* k\omega\ ,
\end{equation}
with $\beta^* = 0.09$. 

As shown in Table~\ref{tbl:inputfeatures}, after combining the invariant and heuristic scalars, the input feature set consists of 33 features. This feature set is relatively large compared to some other data-driven closure investigations \citep{Ling2016, Kaandorp2020}. In machine learning, a large input feature set can lead to overfitting, and a common regularization technique is the elimination of irrelevant input features. However, a brief input feature pruning study during the hyperparameter optimization procedure showed that the validation set performance deteriorated with the elimination of one or more input features from this set. Therefore, it is possible that the availability of even more input features would improve the model performance. A possible method for augmenting the input feature set may be the use of additional nonzero invariant scalars if the flow is three-dimensional. While several of these scalars were found to be zero for two-dimensional flows~\citep{McConkey2022}, the full scalar set may be used as a starting point for fully three-dimensional flows.

\subsection{\label{sec:dataset}Dataset and preprocessing}
\begin{figure}[H]
	\centering
	\includegraphics[width=0.65\linewidth]{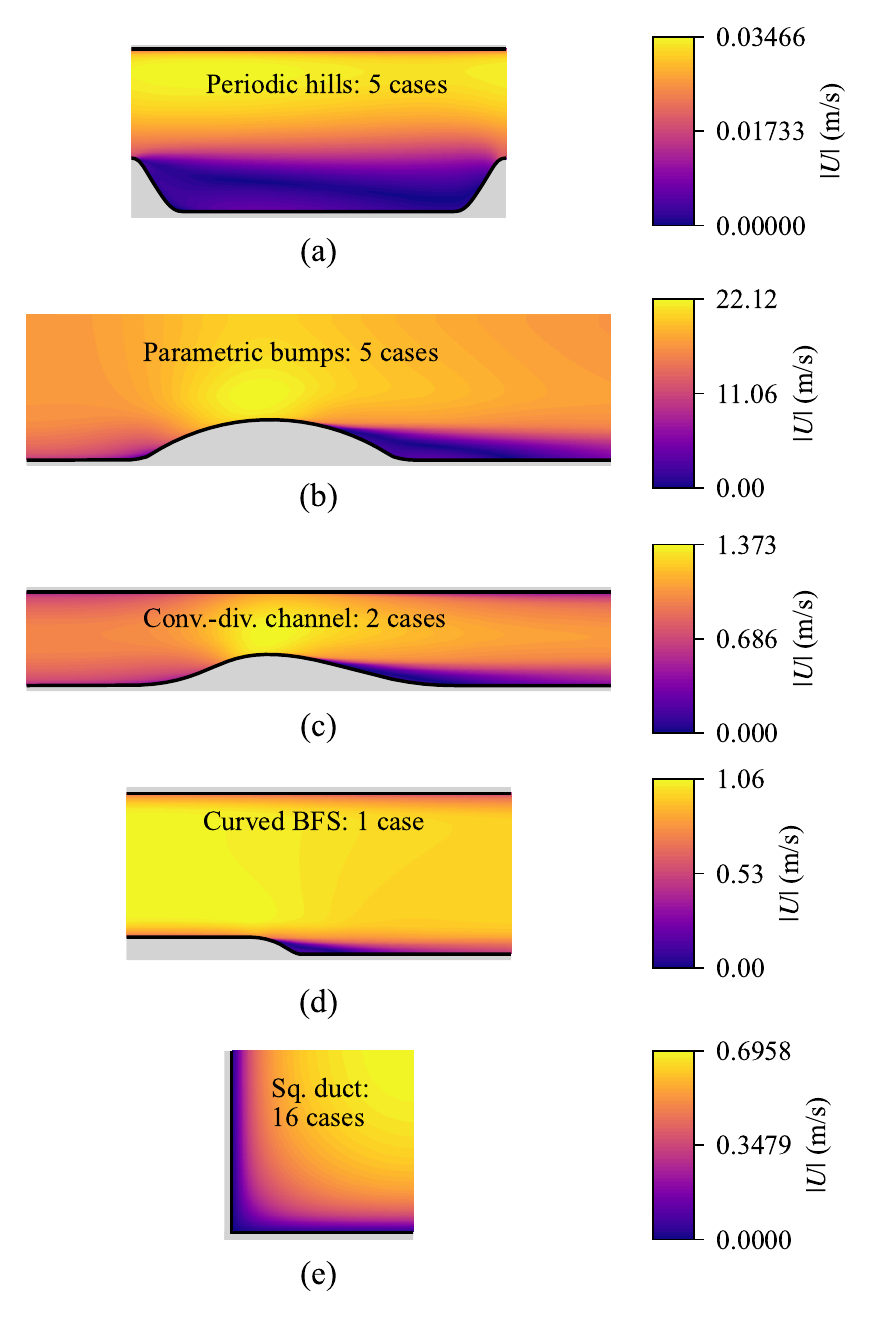}
	\caption{The flows used in the training dataset. The RANS fields for these flows were generated in \cite{McConkeySciDataPaper2021}. The high-fidelity fields were generated for (a) the periodic hills by \cite{Xiao2020}; (b) the parametric bumps by \cite{Matai2019a}; (c) the converging-diverging channel by \cite{Marquillie2008} and \cite{Schiavo2015}; (d) the converging backward-facing step (CBFS) by \cite{Bentaleb2012}; and, (e) the square duct flow by \cite{Pinelli2010}}.
	\label{fig:contour_U}
\end{figure}

A major goal of this study is to investigate the generalizability of various machine learning models for turbulence closure modelling. The training dataset therefore contained several types of flows, included in the dataset compiled by \cite{McConkeySciDataPaper2021}. This dataset consists of several parametrically varying turbulent flows, with collocated RANS and high-fidelity DNS or LES fields. The dataset is designed to be used for generating machine learning mappings between RANS input features and highly accurate closure quantities. Figure~\ref{fig:contour_U} presents an overview of the cases in this curated dataset.

The $k$-$\omega$ SST turbulence model was chosen as the base RANS model~\citep{Menter2003}. This model is popular for simulating separated industrial flows, and provides a reasonably accurate base set of input features. Details of the domain, boundary conditions, mesh (including mesh independence), and numerical schemes for generating the RANS dataset can be found in~\cite{McConkeySciDataPaper2021}.

A total of 895,640 points were available in the base dataset, with all the input features described in Section~\ref{sec:features} pre-calculated. The pre-processing steps included calculating the optimal tensor basis coefficients, removing outliers, and scaling the input features. The results for the optimal tensor basis coefficient calculations are described in Section~\ref{sec:coefficients}. For each point in the dataset, a multiple objective least-squares solver was used to calculate the optimal $g^{(1)}$, $g^{(2)}$, and $g^({3})$ scalar coefficients for the tensor basis expansion of the anisotropy tensor (cf.~Section~\ref{sec:tensorbasis}). 

After calculating the optimal coefficients, several criteria were used to remove outliers. Though the optimization procedure ensured accuracy in the approximation of the anisotropy tensor for the majority of the points, a number of points with a large error in the reconstruction of $b$ still remained. Therefore, the first criterion used for the elimination of outliers was to remove all points corresponding to an absolute error in any component of $b$, $|b_{ij}-b_{ij\theta}|$ greater than 0.05 (for reference, $-2/3 \leq b_{11} \leq 1/3$). The second criterion for outlier removal was based on the magnitude of the basis tensor coefficients themselves. For some locations in the flow, the optimization procedure produced optimal coefficients several orders of magnitude larger than the mean. The thresholds used for the elimination of large $g^{(n)}$ values (outliers) were determined through a trial-and-error process and by inspecting the resulting distributions for $g^{(n)}$ (see Section~\ref{sec:coefficients}). Ultimately, the criteria for this magnitude-based elimination were $g^{(1)}< -0.20$, $|g^{(2)}| > 0.10$, and $|g^{(3)}| > 0.20$. Finally, the square duct case required special treatment in terms of outlier removal. Though the dataset contained points across the entire duct cross section, the duct case is symmetrical about two axes. To avoid including additional data for one flow type that was effectively augmented by reflection, only one quadrant of the square duct cross section was retained. In other words, only one quadrant of the square duct flow was kept in the training dataset in order to eliminate effects of data augmentation through reflection. The error magnitude criteria was also not applied to the square duct coefficients. The square duct case is the only case which contains non-zero $b_{13}$ and $b_{23}$, and therefore all points from the square duct were included due to their unique value. After applying all these criteria for outlier removal, a total of 663,254 points remained in the dataset.

The last step of the pre-processing was scaling of the input features. The scaling method selected was MinMax scaling, where all features are scaled to lie between 0 and 1. Scaling the features to be $O(1)$ helps eliminate issues caused by different orders of magnitudes between each input feature, which is especially important for the neural network model. Although feature scaling has little effect on the results of the tree-based models (viz., random forest and XGBoost), the same scaled input features were used for all the models. The parameters for the MinMax scaler changed with each training set, to consistently ensure that the input features were between 0 and 1.

\subsection{\label{sec:hyperparameter}Training, hyperparameter optimization, and evaluation}
The training procedure used was intertwined with the hyperparameter optimization. To test model generalizability, one hold-out test case was first selected from each flow type (see Table~\ref{tbl:trainingdata}). Then, the remaining cases were used to train and optimize the hyperparameters of each machine learning model. To examine the generalizability of these models in detail, several distinct training datasets were formed. These training datasets are summarized in Table~\ref{tbl:trainingdata}.

The main strategy that was used for forming the various training sets in Table~\ref{tbl:trainingdata} was designed to address the following question: How well does a model trained on one flow generalize to another flow? Despite the important implications of this question for developing a general data-driven turbulence closure model, relatively few studies have examined the generalization performance in this manner. For example, splitting the training data in this way allowed us to test if a model trained purely on periodic hill cases generalized to a parametric bumps case, and vice-versa. In Table~\ref{tbl:trainingdata}, the FULL training set represents the "complete" training set---it contains data from every type of flow except the curved backward facing step (CBFS). As the CBFS data only contain a single case, it represents a hold-out test case for a model trained on the FULL training data set.

After splitting the data into the training and test sets summarized in Table~\ref{tbl:trainingdata}, a hyperparameter optimization procedure was completed. Machine learning models are sensitive to the values of the hyperparameters which define important model configuration variables (e.g., the number of layers in a neural network). Since the optimal hyperparameters depend on the training dataset itself, a hyperparameter optimization procedure must be completed for each new training dataset. 

A grouped four-fold cross-validation (CV) hyperparameter optimization procedure was completed using each training set. This procedure first splits the training dataset into four folds. Each fold is used once as a validation set while the other three folds are used as training sets. The purpose of the validation set is to evaluate the generalization error (variance) after training. A model with a given set of hyperparameters is trained on each training set (three folds), and evaluated on the validation set (the remaining fold). The cross-validation score for a given hyperparameter set is then the average of the four validation set scores. This cross-validation is completed for a number of hyperparameter sets. Then, the optimal hyperparameter set is the one with the best cross-validation score. The purpose of cross-validation is to ensure that the selection of the validation set does not affect the choice of hyperparameters---the average of several validation set choices are used to select the hyperparameters. We recommend using grouped cross-validation for this purpose, to ensure that data points from the same flow case are not present in both the training and validation sets. Grouped cross-validation prevents data leakage and ensures the validation set provides a true measure of the generalization error. For example, in machine learning for medical applications, grouped  cross-validation is used when multiple observations come from the same patient. In this case, four-fold cross validation was selected, as the PHLL4 and BUMP4 training sets contain four cases, which was the minimum number of cases in any of the training sets.

Each machine learning model type (neural network, random forest, and XGBoost) has a different set of hyperparameters that need to be optimized. Full details of the hyperparameter search space are given in~\ref{ap:hyperparameters}. After specifying the search space for each model, the scikit-opt hyperparameter tuning library was used. The hyperparameter tuner was called to complete 100 evaluations of various combinations of hyperparameter values in the search space using Bayesian optimization. Each evaluation in the search space involved the previously described four-fold cross validation procedure. At the end of the 100 evaluations, a model using the optimized hyperparameter set was fit on the entire training dataset. 

The optimized hyperparameters are shown in~\ref{ap:hyperparameters}. For the XGBoost and random forest models, the optimized hyperparameter set produces relatively high-capacity models. While these high-capacity models have a tendency to overfit in some machine learning applications, these hyperparameters were selected for this application using grouped cross-validation. Other tests using models with lower capacity had poorer cross-validation scores and generalization performance on new flows, increasing confidence in the cross-validated hyperparameter search. The high-capacity optimal hyperparameter sets indicate that there is relatively little difference between the validation sets and the training sets. Further work is required to see if this behavior holds for a more diverse 3D flow dataset.

In summary, the training and hyperparameter optimization procedure involved completing 100 iterations of a Bayesian optimizer search within a grouped four-fold cross validation procedure. This procedure was completed for each machine learning model type (see Table~\ref{tbl:modeltypes}), and for each training dataset (see Table~\ref{tbl:trainingdata}). At the end of the hyperparameter optimization procedure, an optimized version of each machine learning model type for each training dataset was then available for evaluation.

An optimized version of each model, together with two simple ensembles were evaluated on the hold-out test datasets. The purpose of this extensive evaluation was to determine the generalizability of each model type on completely new flows (viz., types of flows that were not present in the training dataset), and analyze the effects of including various flow types in the training dataset. Section~\ref{sec:results} presents the results of this comprehensive evaluation.

\subsection{\label{sec:libraries}Libraries and code}
A number of open-source libraries were used for this investigation. Pandas was used as a general data handling library~\citep{pandas}. The optimization procedure for calculating the basis coefficients used the qpsolvers library~\citep{qpsolvers}. For defining model pipelines and evaluating models, scikit-learn was used~\citep{scikit-learn}. The scikit-opt library was used for hyperparameter tuning~\citep{scikit-opt}. Neural network models were constructed using the high-level Keras~\citep{keras} application programming interface (API) for TensorFlow 2, random forest models were created using scikit-learn~\citep{scikit-learn}, and gradient boosting models were created using XGBoost~\citep{xgboostChen2016}. 

All code used for this investigation is available on GitHub as a Jupyter notebook: \url{https://github.com/rmcconke/optimal_tensor_basis}. All data used in this investigation comes from an open-source dataset described by \cite{McConkeySciDataPaper2021}.

\section{\label{sec:results}Results and discussion}

\subsection{\label{sec:coefficients}Coefficients for optimal tensor basis expansion}
\begin{figure}[t]
	\centering
	\includegraphics[width=0.75\linewidth]{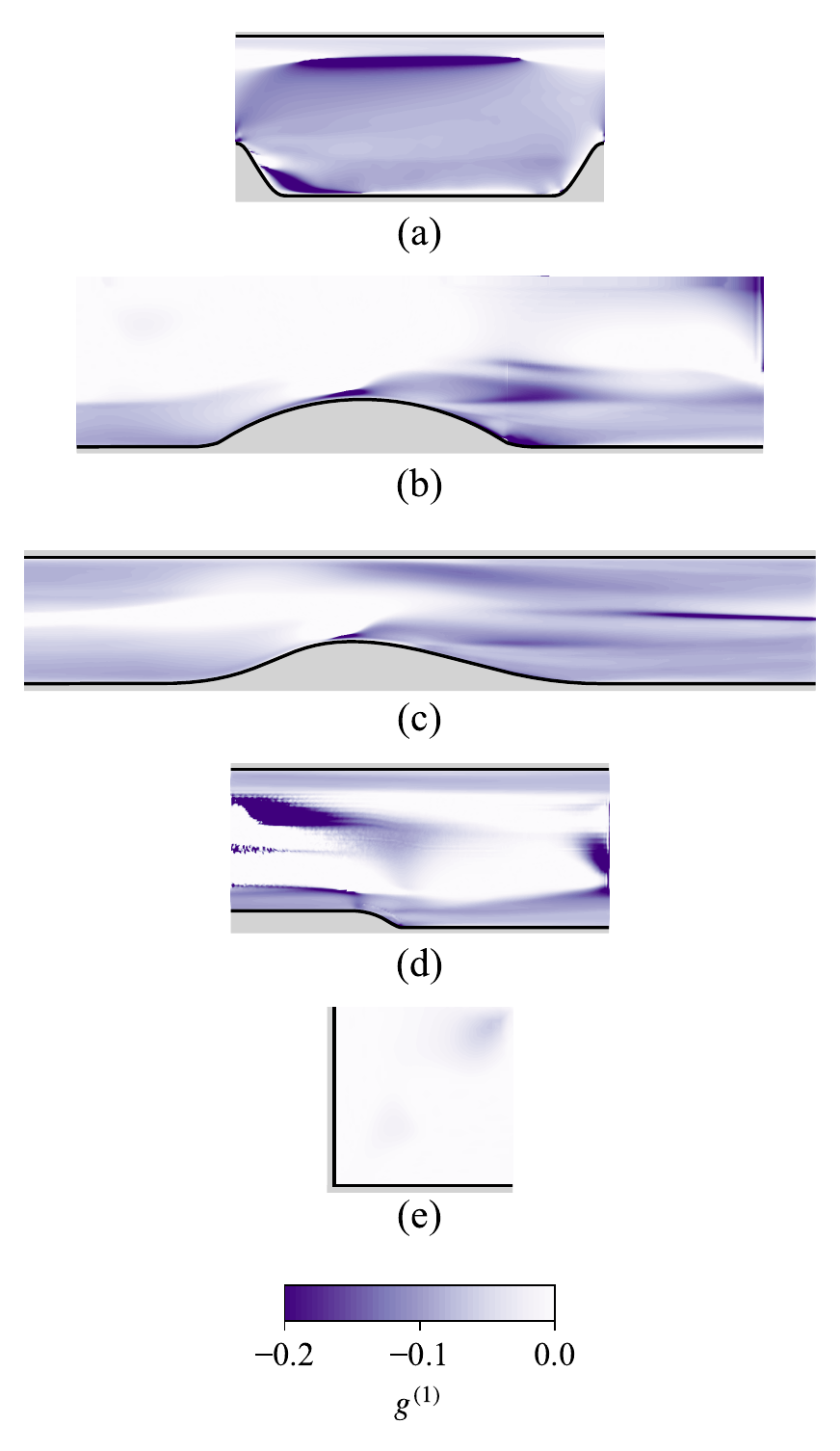}
	\caption{Optimal values for $g^{(1)}$ (cf.~Equation~(\ref{eq:optimization_problem})) for one selected case from each flow type: (a) the periodic hills $\alpha = 0.5$ case; (b) the parametric bump $h=42$ mm case; (c) the converging-diverging channel at a Reynolds number of $\text{Re}=12600$ case; (d) the curved backward-facing step case; and, (e) the square duct at $\text{Re}=1150$ case.}
	\label{fig:contour_g1}
\end{figure}

\begin{figure}[t]
	\centering
	\includegraphics[width=0.75\linewidth]{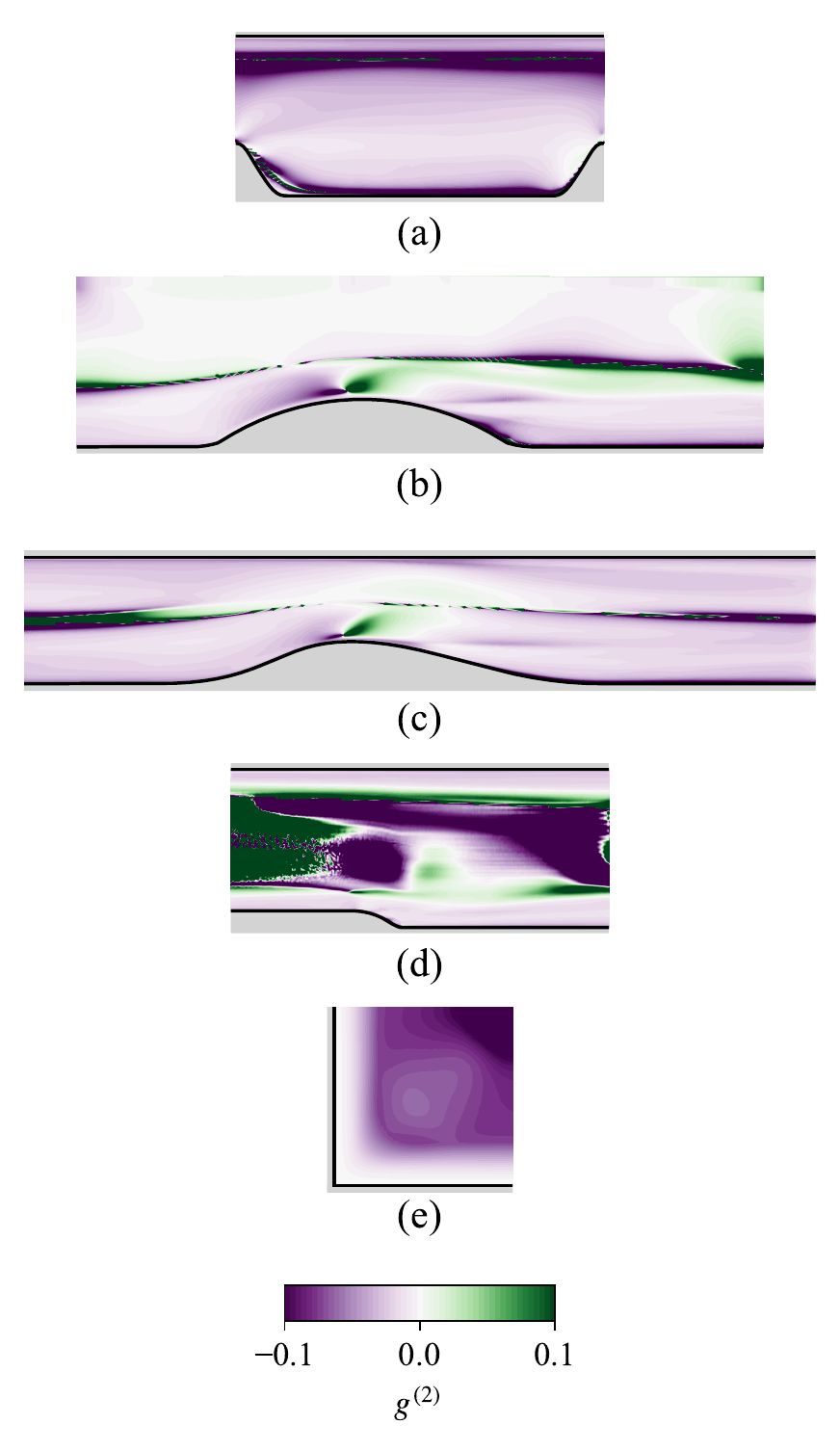}
	\caption{Optimal values for $g^{(2)}$ (cf.~Equation~(\ref{eq:optimization_problem})) for one selected case from each flow type: (a) the periodic hills $\alpha = 0.5$ case; (b) the parametric bump $h=42$ mm case; (c) the converging-diverging channel at a Reynolds number of $\text{Re}=12600$ case; (d) the curved backward-facing step case; and, (e) the square duct at $\text{Re}=1150$ case.}
	\label{fig:contour_g2}
\end{figure}

\begin{figure}[t]
	\centering
	\includegraphics[width=0.75\linewidth]{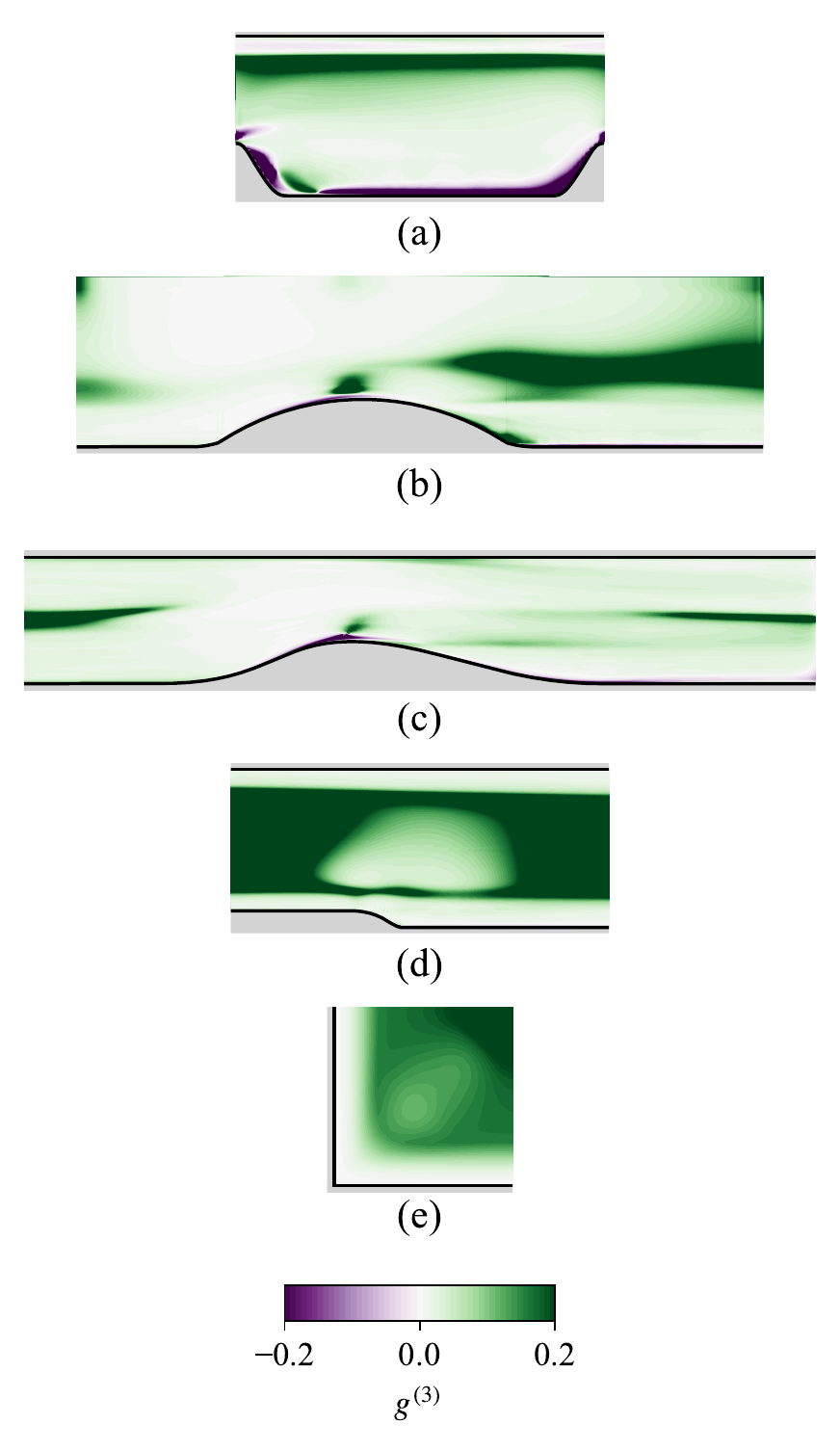}
	\caption{Optimal values for $g^{(3)}$ (cf.~Equation~(\ref{eq:optimization_problem})) for one selected case from each flow type: (a) the periodic hills $\alpha = 0.5$ case; (b) the parametric bump $h=42$ mm case; (c) the converging-diverging channel at a Reynolds number of $\text{Re}=12600$ case; (d) the curved backward-facing step case; and, (e) the square duct at $\text{Re}=1150$ case.}
	\label{fig:contour_g3}
\end{figure}

The optimization problem described in Section~\ref{sec:tensorbasis} was solved using the qpsolvers library~\citep{qpsolvers}. For each data point, optimal values for $g^{(1)}$, $g^{(2)}$, and $g^{(3)}$ were calculated. Figures~\ref{fig:contour_g1} to~\ref{fig:contour_g3} visualize these coefficients for one case from each of the flow types.

Figure~\ref{fig:contour_g1} shows the distinct behavior of $g^{(1)}$ for each flow type. For the periodic hills (Figure~\ref{fig:contour_g1}(a)) $g^{(1)}$ is large in two distinct regions: namely, above the separated region of the flow and near the bottom of the recirculation zone. This same behavior is seen for the bump and the converging-diverging channel cases (Figures~\ref{fig:contour_g1}(b) and (c)). The coefficient $g^{(1)}$ is associated with the mean strain-rate tensor $S$ in the tensor basis expansion of $b$. Therefore, it is not surprising that $g^{(1)}$ is large in near the boundaries of the separated regions of a flow, where the velocity gradients are large. Interestingly, Figure~\ref{fig:contour_g1}(c) shows that near the centerline of the channel upstream of the bump (where the mean strain rate goes to zero), $g^{(1)}$ is small in magnitude. However, along the centerline downstream of the bump, the magnitude of $g^{(1)}$ is large. The Reynolds stress anisotropy is high in this region after the flow accelerates over the bump, leading to the larger values of $g^{(1)}$ seen here. Figure~\ref{fig:contour_g1}(d) shows unusual behavior of $g^{(1)}$ near the top of the domain and along the centerline. As discussed later, unusual behavior of $g^{(2)}$ and $g^{(3)}$ was also observed for the curved backward-facing step (CBFS) case. Although the optimization procedure was able to find a solution for $g^{(n)}$ at all points, large magnitudes for these coefficients were observed for some regions of the CBFS case. Ultimately, these large magnitudes indicate that an accurate approximation of $b$ while minimizing $g^{(n)}$ was not possible in these regions. The likely cause for these instabilities are the extremely small velocity gradients near the channel centerline for the curved backward-facing step. Compared to the other cases, the CBFS features a relatively uniform velocity profile near the channel centerline, where the velocity gradients are small in magnitude. Finally, examining Figure~\ref{fig:contour_g1}(e) shows an unsurprising result---the optimal representation of the anisotropy tensor for the square duct case involves almost no linear component and, therefore, the value of $g^{(1)}$ is small in magnitude. The inability of linear-eddy viscosity models to capture the corner vortices in a square duct is well-known~\citep{Ling2016}.

The coefficients $g^{(2)}$ and $g^{(3)}$ represent the contribution of two basis tensors that are second-order in terms of the mean velocity gradients. For many flows, a purely linear-eddy viscosity model is unable to completely capture the behavior of the anisotropy tensor. Additional nonlinear terms in the tensor basis expansion of $b$ are required to improve this approximation. Figure~\ref{fig:contour_g2} shows the behavior of the optimal values for $g^{(2)}$. Whereas $g^{(1)}$ is strictly negative to ensure stability, no stability constraints are required for the values of $g^{(2)}$. Figures~\ref{fig:contour_g2}(a), (b), and (c) show that $g^{(2)}$ is largest in magnitude near regions of the flow with high velocity magnitude. While the periodic hills case generally exhibits negative values of $g^{(2)}$, both the parametric bump and converging-diverging channel cases exhibit an equal distribution of positive and negative values of $g^{(2)}$. At the peak of the acceleration region in Figures~\ref{fig:contour_g2}(b) and (c), there is a rapid sign change of $g^{(2)}$, with $g^{(2)}$ being negative and positive upstream and downstream of the crest, respectively. Comparing Figure~\ref{fig:contour_g2}(a) to Figures~\ref{fig:contour_g2}(b), (c), (d), and (e), we see that the periodic hills case exhibits a distinctive behavior of $g^{(2)}$ with respect to the other flow cases---these flow cases generally exhibit small $g^{(2)}$ magnitudes near the wall, whereas the periodic hills case exhibits the opposite trend in $g^{(2)}$ near the lower wall. A careful examination of Figure~\ref{fig:contour_g2}(d) indicates that $g^{(2)}$ exhibits an unstable behavior for the CBFS case. Along with the larger magnitudes of $g^{(2)}$ near the inlet centerline, there is a rapid switch in the sign of $g^{(2)}$. In Figure~\ref{fig:contour_g2}(e), we see that the contribution of $T^{(2)}$ is magnified in the region of the square duct corner vortex. This result indicates that a greater contribution of the nonlinear term $T^{(2)}$ is required to capture the anisotropy of the corner vortex---a result that is in agreement with other investigations~\citep{Ling2016}.

Figure~\ref{fig:contour_g3} displays contours of $g^{(3)}$ for the various flow types studied herein. A careful perusal of Figure~\ref{fig:contour_g3} shows that $g^{(2)}$ and $g^{(3)}$ have similar behaviors for some flows, but exhibit distinct behaviors for other flows. For the periodic hills flow (Figure~\ref{fig:contour_g3}(a)), $g^{(3)}$ changes sign within the separated region. The large positive values of $g^{(3)}$ near the top of the domain transition to large negative values near the bottom wall. For both the parametric bump and converging-diverging channel cases (Figures~\ref{fig:contour_g3}(b) and (c)),  $g^{(3)}$ is small in magnitude as the flow accelerates over the obstacle. For these two cases, large positive values of $g^{(3)}$ are observed near the centerline of the domain. Whereas the parametric bump cases impose a zero-gradient (i.e., freestream) boundary condition at the top of the domain, the converging-diverging channel cases impose instead a wall-boundary condition at the top of the domain. This internal/external flow difference may explain the different behaviors of $g^{(3)}$ immediately upstream and downstream of the obstacle. A small negative $g^{(3)}$ region is found immediately upstream of the separation point, along the top of the obstacle. Figure~\ref{fig:contour_g3}(d) shows that similar to $g^{(2)}$, $g^{(3)}$ is large in magnitude near the middle of the domain and smaller magnitudes are observed for the CBFS near the top and bottom walls. However, $g^{(3)}$ does not experience the rapid sign switching observed in $g^{(2)}$ near the inlet. Finally, examining the $g^{(3)}$ field for the square duct case (Figure~\ref{fig:contour_g3}(e)) shows almost identical behavior to that of $g^{(2)}$, but with an opposite sign. The optimal value of $g^{(3)}$ increases near the region of the corner vortex, which agrees with previous results that nonlinear terms in the tensor basis expansion of the anisotropy tensor are needed to capture the corner vortex. 

\begin{figure}[t]
	\centering
	\includegraphics[width=0.75\linewidth]{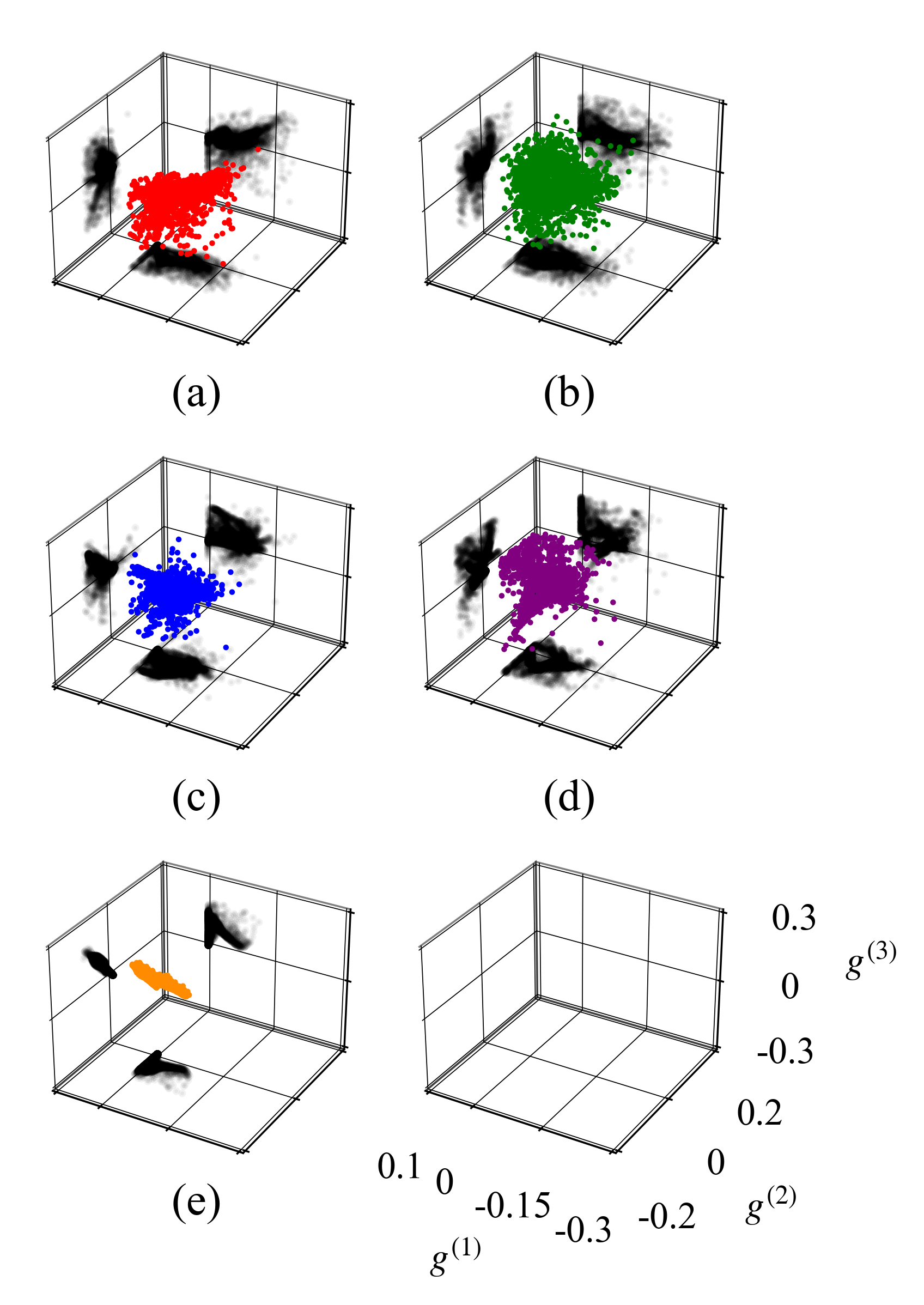}
	\caption{Three-dimensional scatter plot of the optimal $g^{(n)}$ coefficients (cf.~Equation~(\ref{eq:optimization_problem})) for (a) the periodic hills; (b) the parametric bumps; (c) the converging-diverging channels; (d) the curved backward-facing step; and, (e) the square duct.}
	\label{fig:3dscatter}
\end{figure}

Examining the data distribution shape is another important tool for comparing the various flows in this dataset. Figure~\ref{fig:3dscatter} presents a visualization of the data for each flow case, after removal of the outliers (Section~\ref{sec:dataset}). This figure uses each of the $g^{(n)}$ ($n=1,2,3$) scalars as a coordinate to create a three-dimensional point cloud. The projection of this point cloud on the three coordinate planes help visualize the shape of these clouds. From Figure~\ref{fig:3dscatter}, it is immediately clear that the four 2D flows over obstacles have similar data distributions, despite the differences seen in their spatial behavior (cf.~Figures~\ref{fig:contour_g1} to \ref{fig:contour_g3}). The periodic hills, parametric bumps, converging-diverging channel, and curved backward-facing step cases have more diverse optimal coefficients than the square duct cases.

\begin{figure}[t]
	\centering
	\includegraphics[width=0.75\linewidth]{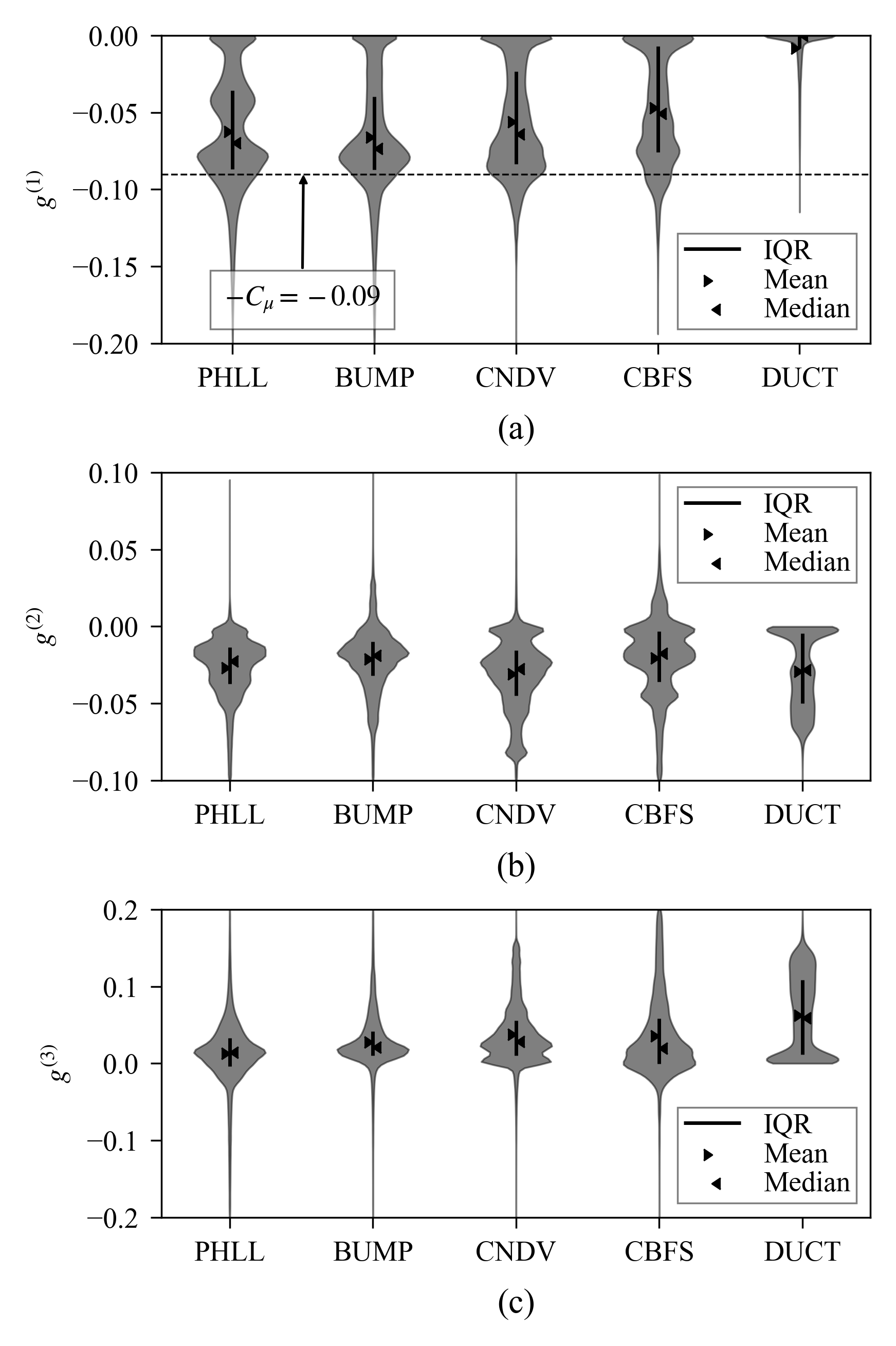}
	\caption{Violin plots showing the distributions of optimal $g^{(n)}$ coefficients: (a) $g^{(1)}$; (b) $g^{(2)}$; and, (c) $g^{(3)}$. Here, IQR denotes the interquartile range (viz., the difference between the 75th and 25th percentiles of the distribution).}
	\label{fig:violin}
\end{figure}

Figure~\ref{fig:violin} displays the distributions of the optimal coefficients for each flow in the form of violin plots. These violin plots, which show a rotated continuous distribution generated using a kernel density estimation methodology, provide an informative comparison between the various flow cases. Figure~\ref{fig:violin} confirms the similarity between the 2D flows over an obstacle---while each distribution displays subtle differences, the overall shape of the distributions are similar. In Figure~\ref{fig:violin}, we see that most of the $g^{(1)}$ distributions exhibit a peak value between about $-0.05$ and $-0.10$. For linear-eddy viscosity models, this coefficient is directly related to the closure coefficient $C_\mu$ as $g^{(1)} = - C_\mu$. This coefficient is applied globally, whereas the optimal $g^{(1)}$ value is a local quantity (viz., it has a dependence on the spatial coordinates). As shown in Figure~\ref{fig:violin}(a), the usual value for this coefficient is approximately equal to the peak value in the optimal $g^{(1)}$ distributions, albeit a little larger in magnitude than the peak value in the $g^{(1)}$ distribution. While $g^{(1)}$ has a direct correspondence with $C_\mu$, no analogous relationships to $g^{(2)}$ and $g^{(3)}$ exists in the classical turbulence closure models, since they represent the contribution to the anisotropy tensor arising from the nonlinear tensor basis terms $T^{(2)}$ and $T^{(3)}$. 

Compared to the other flow cases, the square duct case displays a distinctive behavior in all of the $g^{(n)}$ coefficients. The distribution for $g^{(1)}$ in the square duct case shows that nearly all values are close to zero. While for many points, the optimal values for $g^{(2)}$ and $g^{(3)}$ are close to zero, these coefficients associated with the nonlinear terms in the tensor basis expansion of the anisotropy tensor have much larger magnitude than that of $g^{(1)}$. Clearly, the linear term $T^{(1)}$ embodies virtually no information concerning $b$ for the square duct cases, and the optimization procedure returns an approximation for $b$ consisting almost entirely of the nonlinear contributions from $T^{(2)}$ and $T^{(3)}$. As previously mentioned, a likely reason for this lack of a linear contribution to the anisotropy tensor is due to the failure of a linear-eddy viscosity model to accurately represent the Reynolds stresses associated with this flow.

\subsection{\label{sec:mlresults}Predictions of optimal tensor basis coefficients}
After calculating and visualizing the optimal coefficients $g^{(n)}$ ($n=1,2,3$), it is of interest to determine their predictability using various machine learning models. Furthermore, it is of interest to determine whether a model that was trained to predict the optimal coefficients for one flow would generalize to another flow. As previously noted, these coefficients are used in the present study as a surrogate for predicting the anisotropy tensor (and closing the RANS equations). Therefore, we believe that this generalizability investigation is applicable for many types of data-driven turbulence closure frameworks. 

Table~\ref{tbl:results} shows a matrix of the mean-squared errors (MSE) for various models on various test cases. The models selected are the final versions of each model (viz., the fully-trained machine learning model) using the hyperparameters obtained from the hyperparameter optimization. The MSE is calculated over the three predicted coefficients (viz., $g^{(1)}$, $g^{(2)}$, and $g^{(3)}$).

For every model type, a model was trained on each of the four training datasets. Following this, the trained models were tested on each test set. This exhaustive evaluation allows the generalizability of each model to be determined. As seen in Table~\ref{tbl:results}, all of the models trained (including the naive linear regression model used as a baseline for comparison) generalize well to new cases within the same flow (for example, a different periodic hills geometry). Good generalization (viz., low variance) is identified in this case by having similar CV and test set scores. Across varying flow types such as periodic hills, parametric bumps, and a square duct, Table~\ref{tbl:results} shows that models generalize well within the same flow type. For example, given a training dataset consisting of square duct flows, a machine learning model can be expected to generalize well to new square duct flows. This finding is not new, and has been reported in nearly every data-driven turbulence modelling study---most studies include a generalizability test case within the same flow type. However, a key finding of the present study comes from examining the generalization performance for new flow types. Table~\ref{tbl:results} shows that the MSE increases by roughly one order of magnitude when generalizing to a new flow type. This lack of generalizability holds true across the three machine learning model types and across all four training datasets.

Comparing the results in Table~\ref{tbl:results} for each machine learning model shows that the XGBoost model generally outperforms the random forest and neural network models. The CV and test set scores are generally lower for the XGBoost model, across various training and test datasets. Perhaps the most representative training dataset is the FULL training dataset---this contains every flow in the dataset except the CBFS case. Comparing the performance on the FULL training dataset, we see that the XGBoost model has a similar CV score to the random forest model ($3.035 \times 10^{-4}$ compared to $3.028 \times 10^{-4}$, respectively), and an improved CV score compared to the neural network model. In terms of test set scores, the XGBoost model outperforms the random forest model for the periodic hills, parametric bumps, and square duct test cases, while the random forest performs slightly better on the converging-diverging channel and curved backward-facing step test cases.

Table~\ref{tbl:walltime} compares the computational demands for the hyperparameter optimization procedure. As described in Section~\ref{sec:hyperparameter}, each hyperparameter optimization involved 100 iterations of a Bayesian optimization procedure. Computations were performed on either an AMD Ryzen Threadripper 3970X 32-core central processing unit (CPU), or on an NVIDIA RTX 3090 graphics processing unit (GPU). For the tree-based models, the hyperparameter tuning procedure was completed entirely on a CPU, while the neural network hyperparameter tuning utilized computations on a GPU. Table~\ref{tbl:walltime} shows that the XGBoost model was nearly twice as fast for the hyperparameter optimization as the random forest model, despite a similar or better model performance shown in Table~\ref{tbl:results}. The neural network model tuning was completed faster than the tree-based model tunings, likely due to the use of a GPU to accelerate the calculations. However, the XGBoost model hyperparameter tuning time (viz., 39,285~s) was still similar to that of the neural network model (viz., 30,507~s), despite the use of a CPU versus a GPU. The XGBoost model tuning can be further accelerated by the use of a GPU---this is expected to close or eliminate altogether the differences in times required for hyperparameter tuning of the XGBoost and neural network models.

\begin{figure}[t]
	\centering
	\includegraphics[width=0.75\linewidth]{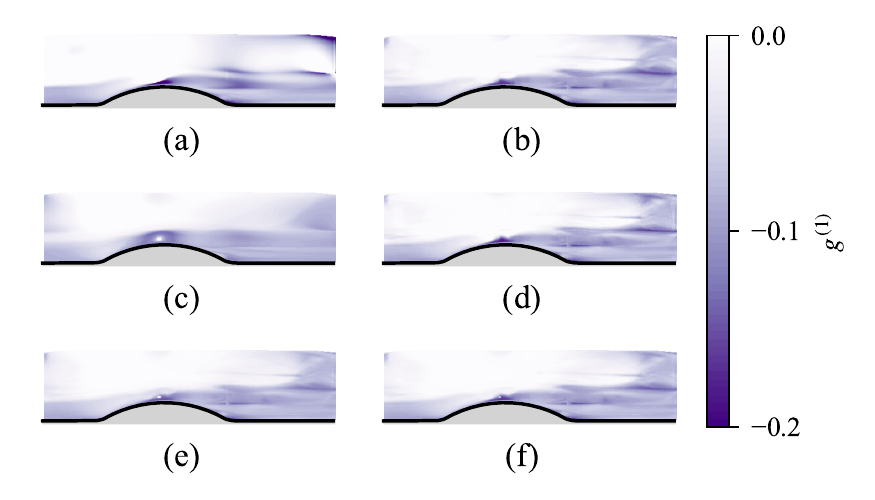}
	\caption{The $g^{(1)}$ fields for the parametric bump test case: (a) true field; (b) random forest model prediction; (c) neural network model prediction; (d) XGBoost model prediction; (e) uniform weighted ensemble prediction; and, (f) CV score weighted ensemble prediction.}
	\label{fig:pred_all_g1_bump}
\end{figure}

\begin{figure}[t]
	\centering
	\includegraphics[width=0.75\linewidth]{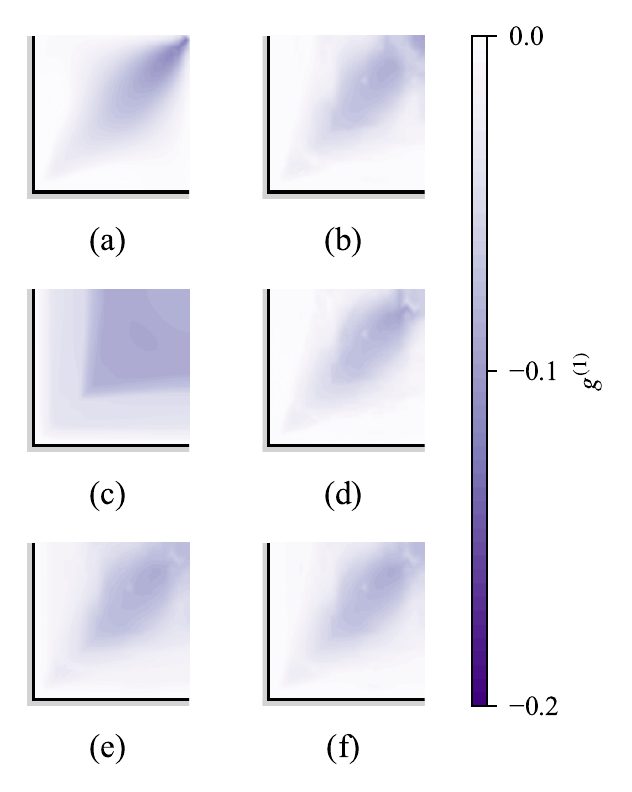}
	\caption{The $g^{(1)}$ fields for the square duct test case: (a) true field; (b) random forest model prediction; (c) neural network model prediction; (d) XGBoost model prediction; (e) uniform weighted ensemble prediction; and, (f) CV score weighted ensemble prediction.}
	\label{fig:pred_all_g1_duct}
\end{figure}

Comparing the three fields for each test case across the five models is not possible within the scope of this paper. A selection is presented here, with further visualizations shown in the supplementary files.  Figures~\ref{fig:pred_all_g1_phll},~\ref{fig:pred_all_g1_bump}, and~\ref{fig:pred_all_g1_duct} show the true and predicted $g^{(1)}$ field for the periodic hill, parametric bump, and square duct test cases, respectively, for the models trained on the combined (FULL) dataset. While the results in Table~\ref{tbl:results} allow easy comparison of MSE magnitudes between models and test cases, visualizing the predicted $g^{(n)}$ fields further facilitates the assessment of the performance of each model. The conclusion obtained from a careful examination of Figures~\ref{fig:pred_all_g1_phll},~\ref{fig:pred_all_g1_bump}, and~\ref{fig:pred_all_g1_duct} is the same: namely, the tree-based models predict the $g^{(n)}$ fields with reasonable accuracy. The neural network model predictions are inferior to those provided by the tree-based models. In particular, from Figure~\ref{fig:pred_all_g1_phll}, the neural network model does not predict the sharp increase in $g^{(1)}$ above the separated region. Furthermore, from a perusal of Figure~\ref{fig:pred_all_g1_bump}, it is seen that the neural network model predictions of the $g^{(n)}$ fields are in close agreement with the associated label (true) fields---however the sharp increase in $g^{(1)}$ near the top of the parametric bump is smoother than that exhibited in the corresponding label field. In contrast, both tree-based models predict this sharp increase in $g^{(1)}$ accurately. Finally, Figure~\ref{fig:pred_all_g1_duct} shows that the neural network model overpredicts $g^{(1)}$ throughout most of the square duct quadrant. The tree-based models accurately capture the shape of this high $g^{(1)}$ region, whereas it is smoother and larger in the neural network model prediction. In Figures~\ref{fig:pred_all_g1_phll},~\ref{fig:pred_all_g1_bump}, and~\ref{fig:pred_all_g1_duct}, the two simple ensembles predict the $g^{(1)}$ field well---which, perhaps, is not too surprising as these two ensemble predictions are obtained simply by averaging over the other three individual predictors. 

Based on the results in Tables~\ref{tbl:results} and~\ref{tbl:walltime}, and the visualizations in Figures~\ref{fig:pred_all_g1_phll} to~\ref{fig:pred_all_g1_duct}, we recommend XGBoost for data-driven turbulence closure modelling involving tabular data. These results are in agreement with recent findings in the broader machine learning community~\citep{Shwartz-Ziv2022}. In terms of MSE magnitude, the XGBoost model performed slightly better than the random forest model for the majority of the test cases, but the difference is minimal. Both tree-based models outperformed the neural network model. The visualizations in Figures~\ref{fig:pred_all_g1_phll} to~\ref{fig:pred_all_g1_duct} also show minimal differences in the fields predicted by the random forest and XGBoost models. However, with the same number of optimizer evaluations, the XGBoost model was faster to optimize---using GPU-based computations is expected to further accelerate this procedure.

Despite poor performance of the neural network shown in Figures~\ref{fig:pred_all_g1_phll},~\ref{fig:pred_all_g1_bump}, and~\ref{fig:pred_all_g1_duct}, neural networks have been used successfully in previous data-driven closure investigations~\citep{Ling2016,McConkey2021c,McConkey2022,Song2019}. Our intention is not to dispute the use of neural networks for data-driven closure modelling, but rather to demonstrate that XGBoost can achieve similar or better performance given the same attention and computational resources. For example, we expect that the neural network model here could be improved by further iterations of the hyperparameter optimizer or exploring various input feature transformations.

\subsubsection{\label{sec:generalizability}Generalizability}
Having selected XGBoost as the recommended model type, we focus now on an analysis of the generalizability performance. In Table~\ref{tbl:results}, we can deduce a rule of thumb that the MSE in $g^{(n)}$ ($n=1,2,3$) is on the same order as the CV score for a new test case within a training flow type, and is approximately within one order of magnitude on a new type of flow. Some improvement in generalizability is expected if more flow types are added to the training dataset. 

\begin{figure}[t]
	\centering
	\includegraphics[width=0.75\linewidth]{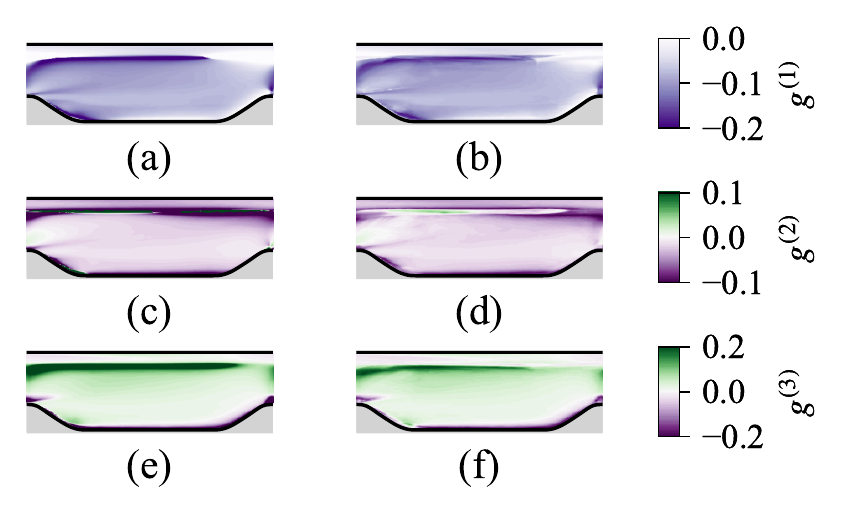}
	\caption{Generalizability test on the periodic hills test case for the XGBoost model. Here, the training dataset is the periodic hills PHLL4 data: (a) true $g^{(1)}$ field; (b) predicted $g^{(1)}$ field; (c) true $g^{(2)}$ field;  (d) predicted $g^{(2)}$ field; (e) true $g^{(3)}$ field; and, (f) predicted $g^{(3)}$ field.}
	\label{fig:trainPHLLtestPHLL}
\end{figure}

Figure~\ref{fig:trainPHLLtestPHLL} shows the hold-out test set performance for the XGBoost model trained on the periodic hills dataset and tested on a periodic hills flow. The predictions for all $g^{(n)}$ component fields are nearly identical to the corresponding label field. This accuracy is also reflected by the low MSE in Table~\ref{tbl:results} for the XGBoost model with the periodic hills training dataset. Figure~\ref{fig:trainPHLLtestPHLL} demonstrates that the model generalizes well to a new case within the same flow type. 

\begin{figure}[t]
	\centering
	\includegraphics[width=0.75\linewidth]{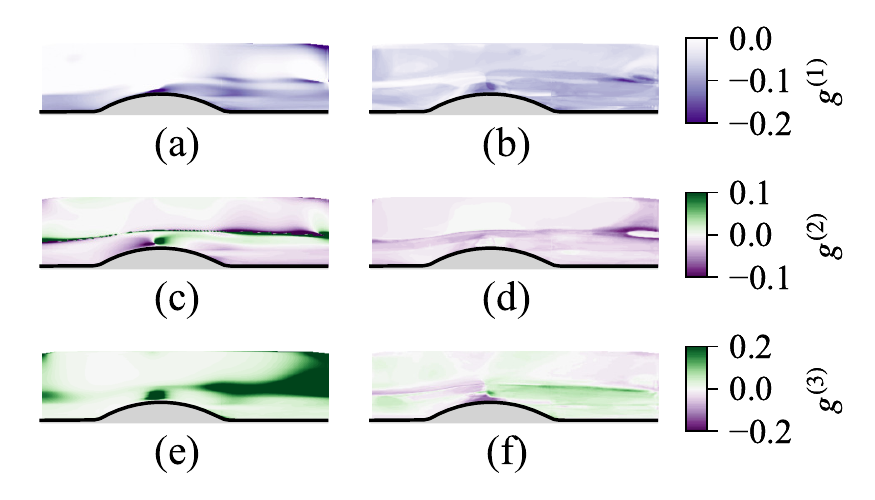}
	\caption{Generalizability test on the parametric bump test case for the XGBoost model. Here, the training dataset is the periodic hills PHLL4 data: (a) true $g^{(1)}$ field; (b) predicted $g^{(1)}$ field; (c) true $g^{(2)}$ field; (d) predicted $g^{(2)}$ field; (e) true $g^{(3)}$ field; and, (f) predicted $g^{(3)}$ field.}
	\label{fig:trainPHLLtestBUMP}
\end{figure}

In order to visualize the generalization gap arising from the application of a data-driven turbulence closure model to a new flow, Figure~\ref{fig:trainPHLLtestBUMP} shows the hold-out test dataset performance for the same periodic hills trained XGBoost model, but using the parametric bump test case. The training dataset for this model only contained the periodic hills data. It is evident that the model clearly does not generalize well to this new type of flow. This is in spite of the fact that these 2D separated flows are similar to one another and have similar $g^{(n)}$ distributions (see Figure~\ref{fig:violin}). Despite these similarities, the periodic hills trained XGBoost model does not generalize well in the prediction of the $g^{(n)}$ fields over a parametric bump. The predictions for the $g^{(1)}$ field are less sharp than the corresponding label field, and there is a distinct sign switching behavior in the $g^{(2)}$ field that is not predicted by the model at all. For the $g^{(3)}$ field, although the sign switching is predicted correctly by the model, the large magnitude of $g^{(3)}$ in the region behind the parametric bump is underpredicted. Table~\ref{tbl:results} shows that the CV score for this model is $1.083 \times 10^{-4}$, whereas the score for the parametric bump test case is $1.116 \times 10^{-3}$ (roughly one order of magnitude worse). In contrast, the score for the periodic hills test case (cf.~Figure~\ref{fig:trainPHLLtestPHLL}) is $3.064 \times 10^{-5}$. A careful perusal of the figures in the supplementary information files further supports this inability of the periodic hills trained XGBoost model to generalize to new types of flow.

\begin{figure}[t]
	\centering
	\includegraphics[width=0.75\linewidth]{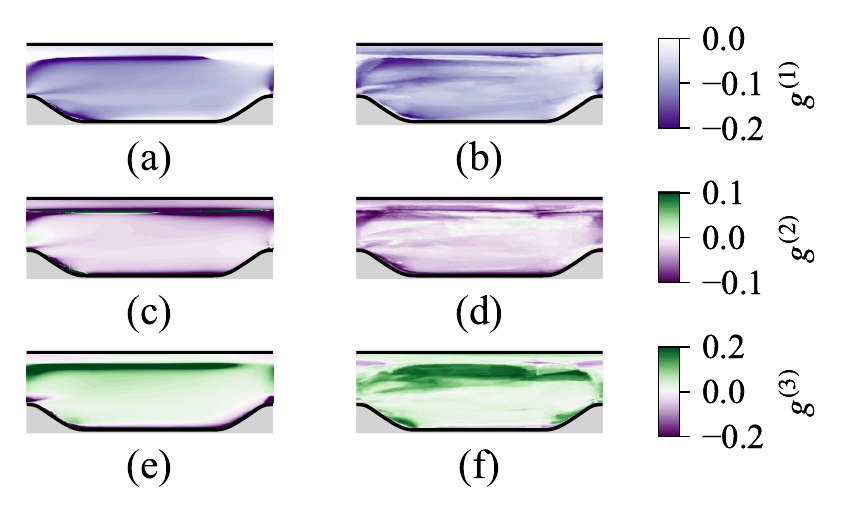}
	\caption{Generalizability test on the periodic hills test case for the XGBoost model. Here, the training dataset is the parametric bump BUMP4 set: (a) true $g^{(1)}$ field; (b) predicted $g^{(1)}$ field; (c) true $g^{(2)}$ field; (d) predicted $g^{(2)}$ field; (e) true $g^{(3)}$ field; and, (f) predicted $g^{(3)}$ field.}
	\label{fig:trainBUMPtestPHLL}
\end{figure}

It is of interest to determine whether the inability to generalize between the periodic hills training dataset and the parametric bump test case would hold for the reverse case. Figure~\ref{fig:trainBUMPtestPHLL} shows the performance of the parametric bump trained XGBoost model on the periodic hills hold-out test case. The generalization performance for the parametric bump training dataset to periodic hills test dataset appears better than that for the reverse case. For the $g^{(1)}$ field, an increase in magnitude above the separation region is predicted. For the $g^{(2)}$ field, the same increase is predicted, in addition to an increase near the lower wall. The $g^{(3)}$ field predictions exhibit similar characteristics. However, the shape and extent of these regions are not nearly as accurate as the predictions provided by the periodic hills trained XGBoost model. 

\begin{figure}[t]
	\centering
	\includegraphics[width=0.75\linewidth]{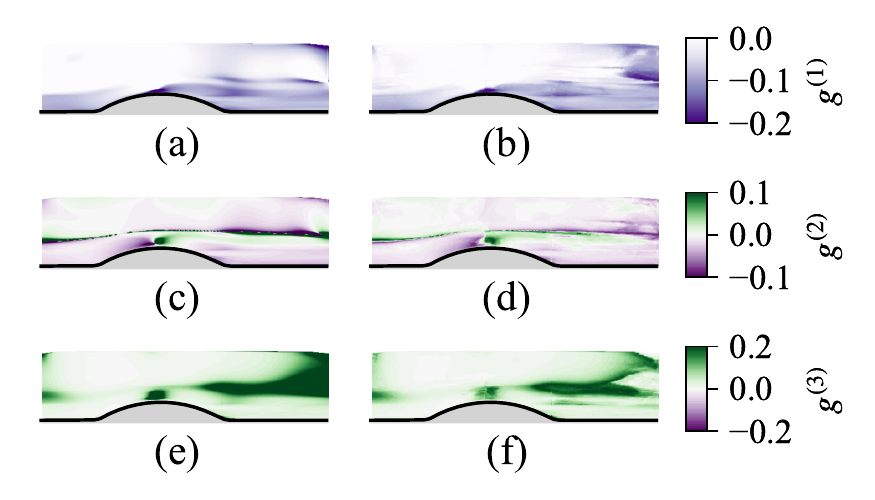}
	\caption{Generalizability test on the parametric bump test case for the XGBoost model. Here, the training dataset is the parametric bump BUMP4 set: (a) true $g^{(1)}$ field; (b) predicted $g^{(1)}$ field; (c) true $g^{(2)}$ field; (d) predicted $g^{(2)}$ field; (e) true $g^{(3)}$ field; and, (f) predicted $g^{(3)}$ field.}
	\label{fig:trainBUMPtestBUMP}
\end{figure}

In Table~\ref{tbl:results}, the CV MSE for this model is $2.484 \times 10^{-4}$, whereas that for the periodic hills test MSE is $1.172 \times 10^{-3}$ (viz., approximately an order of magnitude worse). Despite this poor performance on the periodic hills test case, the model generalizes well to the hold-out parametric bump test case---Figure~\ref{fig:trainBUMPtestBUMP} shows that the predicted and corresponding label fields are nearly identical for the parametric bump trained XGBoost model applied to the hold-out parametric bump test case. The MSE for the parametric bump hold-out test case is $1.262 \times 10^{-4}$ which is on the same order of magnitude as the CV MSE. 

Whereas Figures~\ref{fig:trainPHLLtestPHLL} to~\ref{fig:trainBUMPtestBUMP} demonstrate the generalization gap for two similar 2D flows, the supplementary information files contain several other visualizations (including the generalization performance for the random forest and neural network models). The poor generalization performance shown here extends to all generalizability tests in the present study.

\begin{figure}[t]
	\centering
	\includegraphics[width=0.75\linewidth]{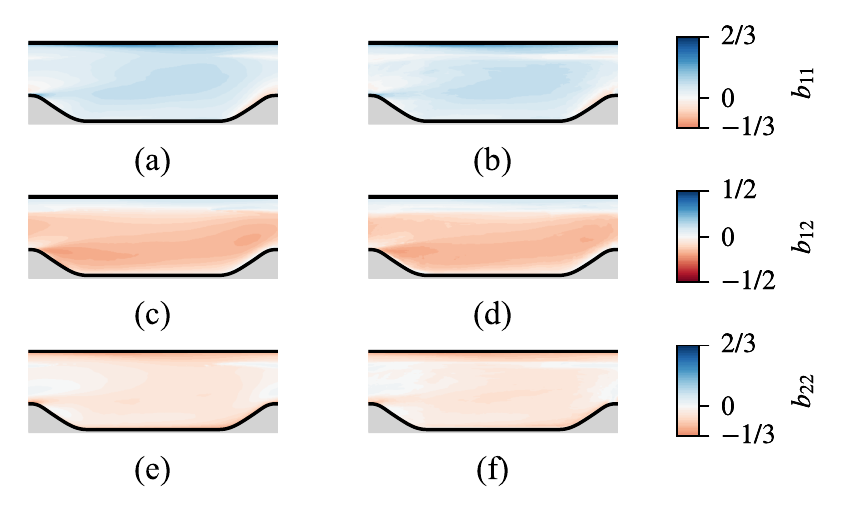}
	\caption{Components of $b$ calculated using the $g^{(n)}$ coefficients from Figure~\ref{fig:trainPHLLtestPHLL}, and the tensors from the base $k$-$\omega$ SST model. Here, the training dataset is the periodic hills PHLL4 set: (a) true $b_{11}$ field; (b) predicted $b_{11}$ field; (c) true $b_{12}$ field; (d) predicted $b_{12}$ field; (e) true $b_{22}$ field; and, (f) predicted $b_{22}$ field.}
	\label{fig:b_phll_phll}
\end{figure}

\begin{figure}[t]
	\centering
	\includegraphics[width=0.75\linewidth]{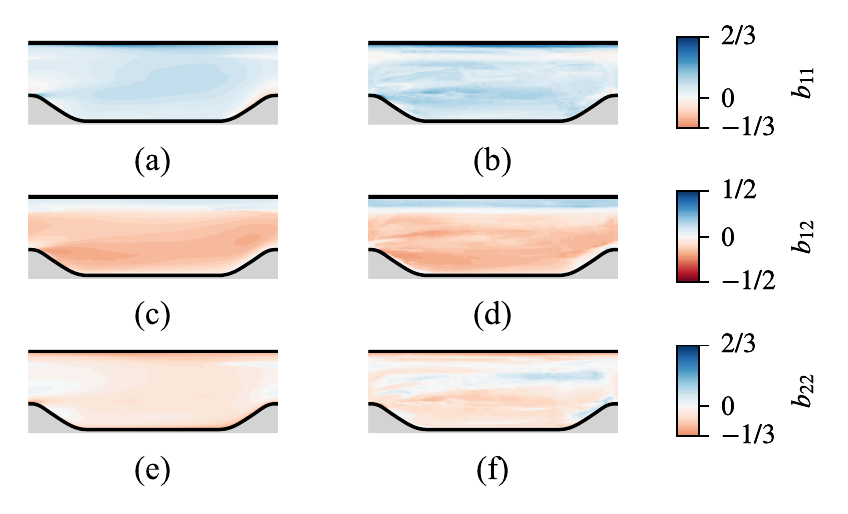}
	\caption{Components of $b$ calculated using the $g^{(n)}$ coefficients from Figure~\ref{fig:trainBUMPtestPHLL}, and the tensors from the base $k$-$\omega$ SST model. Here, the training dataset is the periodic hills BUMP4 set: (a) true $b_{11}$ field; (b) predicted $b_{11}$ field; (c) true $b_{12}$ field; (d) predicted $b_{12}$ field; (e) true $b_{22}$ field; and, (f) predicted $b_{22}$ field.}
	\label{fig:b_bump_phll}
\end{figure}

To further examine the generalizability issue, the anisotropy tensor itself was calculated using the predicted and true $g^{(n)}$ coefficients (Section~\ref{sec:coefficients}). Figures~\ref{fig:b_phll_phll} and~\ref{fig:b_bump_phll} compare the predicted and true values for each component of $b$ for the XGBoost model, for two training sets. While the coefficients $g^{(n)}$ were used in this study as a surrogate for an invariant closure quantity, the anisotropy tensor $b = \sum_{n=1}^3 g^{(n)} T^{(n)}$ is the closure term. Figure~\ref{fig:b_phll_phll} shows that the coefficients from Figure~\ref{fig:trainPHLLtestPHLL} results in an accurately predicted anisotropy tensor. This is expected---accurate predictions of $g^{(n)}$ should lead to accurate computations of $b$. However, Figure~\ref{fig:b_bump_phll} shows that when $g^{(n)}$ are predicted poorly, the prediction of $b$ can be inaccurate. The coefficients in Figure~\ref{fig:trainBUMPtestPHLL} are predicted poorly as a results of the generalization error, which translates to errors in $b$ (Figure~\ref{fig:b_bump_phll}). While $b_{11}$ is predicted with minimal error, $b_{12}$ is overpredicted near the top of the domain. The prediction for $b_{22}$ shows a strong positive value above the right hill, which is not present in the true $b_{22}$ field.

\section{\label{sec:conclusion}Conclusions}

The confluence of machine learning and turbulence closure modelling has led to recent advances in improving the accuracy of RANS simulations. A wide range of closure frameworks and machine learning models have been used to develop sophisticated methods for approximating the Reynolds stress tensor and closing the mean momentum equation in RANS modelling. The goals of this work were to investigate the performance of multiple machine learning model types for use in data-driven turbulence closure modelling, and to test the generalizability of these augmented closure models. 

Towards these goals, we proposed a representative ``invariant closure framework'' that was model-agnostic. This representative closure is an optimal three-term tensor basis expansion of the anisotropy tensor, which forms a complete basis for 2D flows (and a least-squares approximation of a basis for 3D flows). The key closure parameters are three scalar coefficients, one for each basis tensor. The data-driven turbulence closure framework was then formulated as predicting optimal values of these coefficients from a variety of flow input features. We then performed hyperparameter tuning and evaluation of a random forest model, a neural network model, and an XGBoost model. To test generalizability on new flows, we restricted the training dataset to a particular flow type, and then systematically tested the trained model on other flow types. This process was repeated across three different flow types in order to determine the effect of the training dataset on the generalization of the trained turbulence closure model.

Our findings indicate that the XGBoost model achieves a good balance between performance and computational cost. While the raw cross-validation and test dataset performance of the XGBoost model was only slightly better than that of the tuned random forest model, the computational cost was much lower for the XGBoost model tuning process. Both tree-based models outperformed the neural network model. These results are in agreement with findings in the broader machine learning community, which indicate that gradient boosted tree-based ensemble models are often faster and more accurate than neural network predictors for tabular data problems. In the present work, the neural network model was tuned in less wall-clock time than the XGBoost model. However, the XGBoost model training could be accelerated in future studies through the use of GPU computations.

Though the machine learning models generalized well to new parameter variations within the same flow type (e.g., geometry or Reynolds number), the generalization performance on new flows was not nearly as good. These results point towards a slightly more pessimistic outlook for developing a ``universal turbulence model'' using machine learning methods. While a machine learning model can be trained to predict a closure quantity to high accuracy within the same flow type, the model generalization on new types of flows is not reliable. 

The results in the present study were limited to strictly {\it a priori} tests---the predicted closure fields were not injected to update the velocity field. We limited the investigation in this manner in order to allow a more in-depth study of the generalization performance, without introducing other factors such as stability issues after injection. We believe the findings around lack of generalizability will also extend to velocity fields after injection (if they were to converge). If the predicted closure quantities are inaccurate and/or unstable, accuracy of the resulting velocity fields cannot be anticipated.

On a more optimistic note, the generalization gap can be reduced by incorporating a richer and broader set of flows in the training dataset. In the present work, a model trained on a variety of flows was able to generalize well within new variations of all flows in the training dataset. As the wide availability of open-source datasets in turbulence increases, the ability to incorporate a broader set of flows in a training dataset also increases. While the dataset used in the present work is perhaps the most diverse single-format dataset investigated to date, it only consists of five different 2D flows. We therefore recommend that more DNS and LES datasets be released publicly in order to accelerate data-driven modelling.

Furthermore, the lack of generalizability shown in the present work is not surprising. For decades, turbulence modellers have battled with the ``no free lunch theorem''---a sophisticated model tuned for one flow often does not generalize well to other flows. Despite this, models highly suited for particular flow types are still widely used. For many industrial applications, a universal turbulence model (if such a model even exists) is not required---a tailored turbulence model for the particular application is what is needed. As an example, the Spalart-Allmaras one-equation model has received widespread use for external aerodynamics simulations~\citep{Spalart1992}, where it performs well despite lack of generalizability to some other flows. This suggests that many industrial users prefer an accurate purpose-built turbulence model over a less accurate ``universal'' turbulence model. Several investigations have already developed problem-specific data-driven closures with good success~\citep{Nikolaou2020,Maulik2021b}

We believe the use in machine learning methods for turbulence modelling simply offers a new set of tools and paradigms for developing turbulence closures that are highly accurate within a specific flow type. Nearly all investigations into data-driven RANS closures demonstrate excellent performance, and in many cases the resulting velocity fields can be almost identical to DNS results. Machine learning methods offer industrial users a way to leverage large (internal or public) datasets to develop accurate models for their own use cases. For example, an automotive user can develop a model which is highly accurate for predicting separated flow over vehicles, but that does not generalize well for flows through rotating turbomachinery. Another user focused on rotating flows can use machine learning to develop a well-tuned turbulence closure model for rotating flows that does not necessarily generalize well to aerodynamic flows. Whereas hand-tuning nonlinear or sophisticated RANS models in the past was expensive or difficult, machine learning methods offer a data-driven alternative to develop turbulence closure models for specific use cases.

\setcounter{section}{0}
\renewcommand{\thesection}{Appendix \Alph{section}}
\section{\label{ap:hyperparameters} Hyperparameter Search Space}

Table~\ref{tbl:hyperparameters} shows the search space and tuning results for the hyperparameter optimization.

\section*{Acknowledgments}
R.M. was supported by the Ontario Graduate Scholarship (OGS) program and the Natural Sciences and Engineering Research Council of Canada (NSERC). The computational resources for this work were supported by the Tyler Lewis Clean Energy Research Foundation (TLCERF) and the Shared Hierarchical Academic Research Computing Network (SHARCNET).

The authors wish to thank the anonymous reviewers for their careful review of our manuscript and helpful suggestions.

\section*{Disclosure statement}
The authors report there are no competing interests to declare.

\section*{Data Availability}
All data used in this study are from an open-source turbulence modelling dataset on Kaggle: \url{https://doi.org/10.34740/kaggle/dsv/2637500}. All code used in this investigation is available on github: \url{https://github.com/rmcconke/optimal_tensor_basis}.

\bibliography{references}

\newpage
\section*{Tables}
\begin{table}[H]
	\caption{Types of machine learning methodologies for turbulence closure modelling used in the present study.}\label{tbl:modeltypes}
	\bigskip
	\centering
	\begin{tabular}{cc}
		\hline
		Model          & Type                             \\ \hline
		Neural network & Fully connected, feedforward MLP \\
		Random forest  & Tree-based ensemble model        \\
		XGBoost        & Gradient boosted tree ensemble   \\ \hline
	\end{tabular}
\end{table}

\begin{table}[H]
	\caption{Input feature set used for all machine learning models. The first and second matrix invariants of a matrix $A$ are defined as $I_1(A)\equiv \text{tr}(A)$ and $I_2 \equiv \tfrac{1}{2}[(\text{tr}(A))^2 - \text{tr}(A^2)]$, respectively. Expressions for $q_1$, $q_2$, and $q_3$ are given in Equations~(\ref{eq:q1}), (\ref{eq:q2}), and (\ref{eq:q3}), respectively.} \label{tbl:inputfeatures}
	\bigskip
	\centering
	\begin{tabular}{ccc}
		\hline
		$I_1$        & $I_2$      & Scalars \\ \hline
		$S^2$        & $S^2$      & $q_1$ \\
		$R^2$        & $S^3$      & $q_2$ \\
		$A_p^2$      & $R^2$      & $q_3$   \\
		$A_k^2$      & $A_p^2$    &  --       \\
		$R^2S^2$     & $A_k^2$    &  --       \\
		$A_p^2S$     & $R^2S$     &  --       \\
		$A_p^2S^2$   & $R^2S^2$   &  --       \\
		$A_k^2 S$    & $R^2SRS^2$ &  --       \\
		$A_k^2S^2$   & $A_pA_k$   &  --       \\
		$A_pA_k$     &    --      &  --       \\
		$A_pRS$      &    --       & --         \\
		$A_p^2SRS^2$ &    --       & --        \\
		$A_k^2RS$    &    --       & --        \\
		$A_kSRS^2$   &    --       & --        \\
		$A_pA_kS$    &    --       & --        \\
		$A_pA_kS^2$  &    --       & --        \\
		$RA_pA_k$    &    --       & --        \\
		$RA_pA_kS$   &    --       & --        \\
		$RA_kA_pS$   &    --       & --        \\
		$RA_pA_kS^2$ &    --       & --        \\
		$RA_kA_pS^2$ &    --       & --        \\ \hline
	\end{tabular}
\end{table}

\begin{table}[H]
	\caption{Training and test datasets: periodic hill PHLL; parametric bump BUMP; square duct DUCT; converging-diverging channel CNDV; and, curved backward-facing step CBFS.}\label{tbl:trainingdata}
	\bigskip
	\centering
\begin{tabular}{ccc}
\hline
                                                                                              & Name                                   & Cases                                                                                                                                                                                                                                  \\ \hline
\multicolumn{1}{c|}{\multirow{4}{*}{\begin{tabular}[c]{@{}c@{}}Training\\ sets\end{tabular}}} & PHLL4                                  & \begin{tabular}[c]{@{}c@{}}PHLL\_case\_0p5, PHLL\_case\_0p8, PHLL\_case\_1p0,\\  PHLL\_case\_1p5\end{tabular}                                                                                                                          \\ \cline{2-3} 
\multicolumn{1}{c|}{}                                                                         & BUMP4                                  & \begin{tabular}[c]{@{}c@{}}BUMP\_h20, BUMP\_h26, BUMP\_h31, \\ BUMP\_h42\end{tabular}                                                                                                                                                  \\ \cline{2-3} 
\multicolumn{1}{c|}{}                                                                         & DUCT15                                 & \begin{tabular}[c]{@{}c@{}}DUCT\_1100, DUCT\_1150, DUCT\_1250, \\ DUCT\_1300, DUCT\_1350, DUCT\_1400, \\ DUCT\_1500, DUCT\_1600, DUCT\_1800, \\ DUCT\_2205, DUCT\_2400, DUCT\_2600, \\ DUCT\_2900, DUCT\_3200, DUCT\_3500\end{tabular} \\ \cline{2-3} 
\multicolumn{1}{c|}{}                                                                         & FULL                                   & \begin{tabular}[c]{@{}c@{}}PHLL4, BUMP4, DUCT15, \\ CNDV\_12600\end{tabular}                                                                                                                                                           \\ \hline
\multicolumn{1}{c|}{\multirow{5}{*}{\begin{tabular}[c]{@{}c@{}}Testing\\ sets\end{tabular}}}  & Periodic hill test case                & PHLL\_case\_1p2                                                                                                                                                                                                                        \\ \cline{2-3} 
\multicolumn{1}{c|}{}                                                                         & Bump test case                         & BUMP\_h38                                                                                                                                                                                                                              \\ \cline{2-3} 
\multicolumn{1}{c|}{}                                                                         & Square duct test case                  & DUCT\_2000                                                                                                                                                                                                                             \\ \cline{2-3} 
\multicolumn{1}{c|}{}                                                                         & Converging-diverging channel test case & CNDV\_20580                                                                                                                                                                                                                            \\ \cline{2-3} 
\multicolumn{1}{c|}{}                                                                         & Curved backward facing step test case  & CBFS\_13700                                                                                                                                                                                                                            \\ \hline
\end{tabular}
\end{table}

\begin{table}[H]
    \setlength\tabcolsep{4pt}
	\caption{Mean squared error (MSE) for various training and test datasets for each model type.}\label{tbl:results}
	\bigskip
	\centering
	\begin{tabular}{cccccccc}
		\hline
		&              & \multicolumn{6}{c}{Mean squared error}                                \\ \hline
		Model                                                                                      & Training set & CV score  & PHLL      & BUMP      & DUCT      & CNDV20580 & CBFS      \\ \hline
		\multicolumn{1}{c|}{\multirow{4}{*}{Linear regression}}                                    & PHLL4        & 7.402E-04 & 6.663E-04 & 1.063E+04 & 2.001E-03 & 5.004E+00 & 4.399E-03 \\
		\multicolumn{1}{c|}{}                                                                      & BUMP4        & 1.310E+02 & 1.015E-03 & 1.035E-03 & 1.200E-03 & 1.157E-03 & 1.409E-03 \\
		\multicolumn{1}{c|}{}                                                                      & DUCT15       & 2.272E-04 & 7.152E+7  & 9.328E+11 & 1.734E-04 & 6.833E+10 & 1.441E+08 \\
		\multicolumn{1}{c|}{}                                                                      & FULL         & 1.310E+02 & 1.015E-03 & 1.035E-03 & 1.200E-03 & 1.157E-03 & 1.409E-03 \\ \hline
		\multicolumn{1}{c|}{\multirow{4}{*}{Neural network}}                                       & PHLL4        & 6.365E-04 & 3.056E-04 & 3.046E+00 & 1.532E-03 & 2.062E-02 & 2.529E-03 \\
		\multicolumn{1}{c|}{}                                                                      & BUMP4        & 5.168E-04 & 1.496E-01 & 4.714E-04 & 1.001E-03 & 2.616E+01 & 6.430E-04 \\
		\multicolumn{1}{c|}{}                                                                      & DUCT15       & 4.130E-05 & 1.568E+06 & 4.848E+06 & 2.746E-05 & 8.671E+05 & 2.349E+06 \\
		\multicolumn{1}{c|}{}                                                                      & FULL         & 7.541E-04 & 1.088E-03 & 9.735E-04 & 7.337E-04 & 1.047E-03 & 9.081E-04 \\ \hline
		\multicolumn{1}{c|}{\multirow{4}{*}{Random forest}}                                        & PHLL4        & 1.296E-04 & 3.642E-05 & 1.154E-03 & 1.623E-03 & 1.506E-03 & 2.514E-03 \\
		\multicolumn{1}{c|}{}                                                                      & BUMP4        & 2.760E-04 & 1.172E-03 & 1.315E-04 & 9.068E-04 & 1.036E-03 & 8.232E-04 \\
		\multicolumn{1}{c|}{}                                                                      & DUCT15       & 1.215E-05 & 3.621E-03 & 3.238E-03 & 9.796E-06 & 1.407E-03 & 2.063E-03 \\
		\multicolumn{1}{c|}{}                                                                      & FULL         & 3.028E-04 & 9.667E-05 & 1.608E-04 & 1.951E-05 & 8.363E-04 & 1.016E-03 \\ \hline
		\multicolumn{1}{c|}{\multirow{4}{*}{XGBoost}}                                              & PHLL4        & 1.083E-04 & 3.064E-05 & 1.116E-03 & 1.495E-03 & 1.536E-03 & 1.865E-03 \\
		\multicolumn{1}{c|}{}                                                                      & BUMP4        & 2.484E-04 & 1.384E-03 & 1.262E-04 & 8.000E-04 & 1.083E-03 & 8.404E-04 \\
		\multicolumn{1}{c|}{}                                                                      & DUCT15       & 1.140E-05 & 3.434E-03 & 3.057E-03 & 6.729E-06 & 1.456E-03 & 2.545E-03 \\
		\multicolumn{1}{c|}{}                                                                      & FULL         & 3.035E-04 & 3.330E-05 & 1.360E-04 & 9.413E-06 & 8.718E-04 & 1.164E-03 \\ \hline
		\multicolumn{1}{c|}{\begin{tabular}[c]{@{}c@{}}Equal weight\\ ensemble\end{tabular}}       & FULL         &           & 1.907E-04 & 2.528E-04 & 9.122E-05 & 8.663E-04 & 9.216E-04 \\ \hline
		\multicolumn{1}{c|}{\begin{tabular}[c]{@{}c@{}}CV score \\ weighted ensemble\end{tabular}} & FULL         &           & 7.506E-05 & 1.603E-04 & 2.092E-05 & 8.486E-04 & 1.013E-03 \\ \hline
	\end{tabular}
\end{table}

\begin{table}[H]
	\caption{Wall-clock time taken for the hyperparameter tuning procedure on the FULL training dataset. All central processing unit (CPU) based calculations (random forest and XGBoost) used an identical machine and number of cores, and the neural network calculations used a graphics processing unit (GPU). }\label{tbl:walltime}
	\bigskip
	\centering
	\begin{tabular}{ccc}
		\hline
		Model type     & Resource type & Wall time (s) \\ \hline
		Neural network & GPU           & 30507         \\
		Random forest  & CPU           & 70603         \\
		XGBoost        & CPU           & 39285         \\ \hline
	\end{tabular}
\end{table}

\setcounter{table}{0}
\renewcommand{\thetable}{A\arabic{table}}
\begin{table}[H]
	\caption{Hyperparameter tuning search space and results for each model. Hyperparameter tuning used 100 iterations of a Bayesian optimizer~\citep{scikit-opt}. Shown here are the optimal values on the FULL training set (see Table~\ref{tbl:trainingdata}). However, this procedure was completed to search for an optimized set of hyperparameters for each training dataset. The supplementary information includes the results for each training dataset.}\label{tbl:hyperparameters}
	\bigskip
	\centering
	\begin{tabular}{cccc}
		\hline
		Model                                                                                          & Hyperparameter                             & Search space         & \begin{tabular}[c]{@{}c@{}}Optimal value\\ (FULL training set)\end{tabular} \\ \hline
		\multicolumn{1}{c|}{\multirow{6}{*}{\begin{tabular}[c]{@{}c@{}}Neural\\ network\end{tabular}}} & Activation function                        & ReLU, SELU, or ELU   & ReLU                                                                        \\
		\multicolumn{1}{c|}{}                                                                          & Number of hidden layers                    & {[}3, 20{]}          & 4                                                                           \\
		\multicolumn{1}{c|}{}                                                                          & Number of neurons per layer                & {[}30, 800{]}        & 362                                                                         \\
		\multicolumn{1}{c|}{}                                                                          & L1 regularization parameter                & $[10^{-8}, 10^{-1}]$ & $2.32(10)^{-6}$                                                             \\
		\multicolumn{1}{c|}{}                                                                          & L2 regularization parameter                & $[10^{-8}, 10^{-1}]$ & $10^{-8}$                                                                   \\
		\multicolumn{1}{c|}{}                                                                          & Learning rate                              & $[10^{-7}, 10^{-3}]$ & $8.29(10)^{-4}$                                                             \\ \hline
		\multicolumn{1}{c|}{\multirow{6}{*}{\begin{tabular}[c]{@{}c@{}}Random\\ forest\end{tabular}}}  & Number of estimators                       & {[}100, 1000{]}      & 889                                                                         \\
		\multicolumn{1}{c|}{}                                                                          & Max depth                                  & {[}5, 50{]}          & 28                                                                          \\
		\multicolumn{1}{c|}{}                                                                          & Minimum samples to split internal node     & $[10^{-5}, 10^{-1}]$ & $1.0110)^{-5}$                                                              \\
		\multicolumn{1}{c|}{}                                                                          & Minimum samples required at leaf node      & $[10^{-5}, 10^{-1}]$ & $4.04(10)^{-5}$                                                             \\
		\multicolumn{1}{c|}{}                                                                          & Maximum fraction of features for estimator & {[}0.5, 1.0{]}       & 0.5                                                                         \\
		\multicolumn{1}{c|}{}                                                                          & Maximum number of samples for estimator    & {[}0.5, 1.0{]}       & 0.520                                                                       \\ \hline
		\multicolumn{1}{c|}{\multirow{5}{*}{XGBoost}}                                                  & Number of estimators                       & {[}100, 1000{]}      & 887                                                                         \\
		\multicolumn{1}{c|}{}                                                                          & Max depth                                  & {[}5, 50{]}          & 37                                                                          \\
		\multicolumn{1}{c|}{}                                                                          & Learning rate                              & $[10^{-2},1]$        & $1.30(10)^{-2}$                                                             \\
		\multicolumn{1}{c|}{}                                                                          & Maximum fraction of features for estimator & {[}0.5, 1.0{]}       & 0.562                                                                       \\
		\multicolumn{1}{c|}{}                                                                          & Maximum number of samples for estimator    & {[}0.5, 1.0{]}       & 0.788                                                                       \\ \hline
	\end{tabular}
\end{table}

\newpage
\section*{Figures}

\begin{figure}[t]
	\centering
	\includegraphics[width=0.75\linewidth]{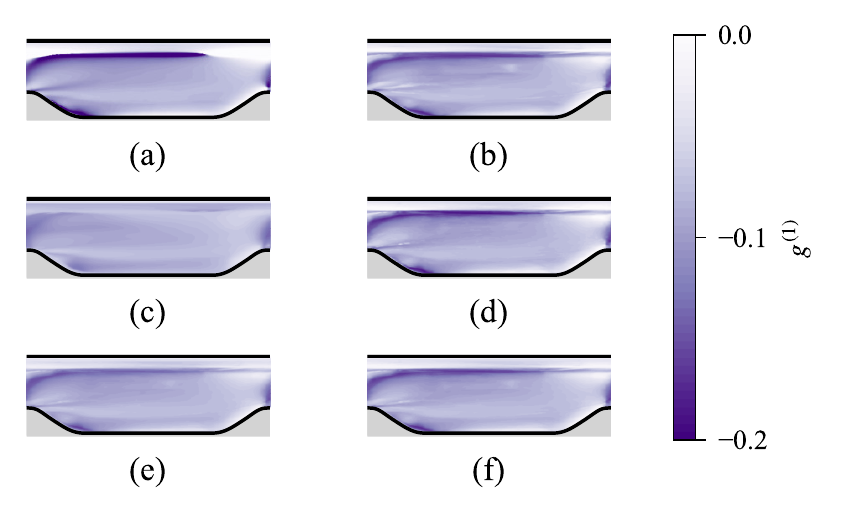}
	\caption{The $g^{(1)}$ fields for the periodic hills test case: (a) true field; (b) random forest model prediction; (c) neural network model prediction; (d) XGBoost model prediction; (e) uniform weighted ensemble prediction; and, (f) CV score weighted ensemble prediction.}
	\label{fig:pred_all_g1_phll}
\end{figure}

\end{document}